\documentclass[referee, sn-standardnature]{sn-jnl}

\usepackage[T1]{fontenc}
\usepackage[version=3]{mhchem}
\usepackage{hyperref}

\newcommand{\DDEsolv}{$\Delta \Delta E_\mathrm{solv}$}
\newcommand{\tcm}{$\eta_\text{\,TCM}$}

\jyear{2023}%

\begin{document}

\title[B-Doped Graphene Solvation]{On the challenge of obtaining an accurate solvation energy estimate in simulations of electrocatalysis}




\author*[1,2]{\fnm{Björn} \sur{Kirchhoff}}\email{bjoern.kirchhoff@uni-ulm.de}

\author[2]{\fnm{Elvar Ö.} \sur{Jónsson}}

\author[1,3,4]{\fnm{Timo} \sur{Jacob}}

\author[2]{\fnm{Hannes} \sur{Jónsson}}

\affil[1]{\orgdiv{Institute of Electrochemistry}, \orgname{Ulm University}, \orgaddress{\street{Albert-Einstein-Allee 47}, \city{Ulm}, \postcode{89081}, \country{Germany}}}

\affil[2]{\orgdiv{Science Institute and Faculty of Physical Sciences}, \orgname{University of Iceland}, \orgaddress{\street{VR-III, Hjarðarhagi 2}, \city{Reykjavík}, \postcode{107}, \country{Iceland}}}

\affil[3]{\orgdiv{Helmholtz-Institute Ulm (HIU) Electrochemical Energy Storage}, \orgaddress{\street{Helmholtz-Straße 16}, \city{Ulm}, \postcode{89081}, \country{Germany}}}

\affil[4]{\orgname{Karlsruhe Institute of Technology (KIT)}, \orgaddress{\street{P.O. Box 3640}, \city{Karlsruhe}, \postcode{76021}, \country{Germany}}}

\abstract{
The effect of solvent on the free energy of reaction intermediates adsorbed on electrocatalyst surfaces can significantly change the thermochemical overpotential, but accurate calculations of this are challenging. Here, we present computational estimates of the solvation energy for reaction intermediates in oxygen reduction reaction (ORR) on a B-doped graphene (BG) model system where the overpotential is found to reduce by up to 0.6 V due to solvation. BG is experimentally reported to be an active ORR catalyst but recent computational estimates using state-of-the-art hybrid density functionals in the absence of solvation effects have indicated low activity. To test whether the inclusion of explicit solvation can bring the calculated activity estimates closer to the experimental reports, up to 4 layers of water molecules are included in the simulations reported here.
The calculations are based on classical molecular dynamics and local minimization of energy using atomic forces evaluated from electron density functional theory.
Data sets are obtained from
regular and coarse-grained dynamics, as well as local minimization of structures resampled from dynamics simulations. The results differ greatly depending on the method used and the solvation energy estimates and are deemed untrustworthy. It is concluded that a significantly larger number of water molecules is required to obtain converged results for the solvation energy. As the present system includes up to 139 atoms, it
already strains the limits of computational feasibility, so this points to the need for a hybrid simulation approach where efficient simulations of much larger number of solvent molecules is carried out using a lower level of theory while retaining the higher level of theory for the reacting molecules as well as their near neighbors and the catalyst. The results reported here provide a word of caution to the computational catalysis community:  activity predictions can be inaccurate if too few solvent molecules are included in the calculations.
}


\maketitle

\section{Introduction}\label{s:introduction}

The replacement of costly and rare precious metals with cheaper and more abundant elements in catalysts, for example in the oxygen reduction reaction (ORR)
in fuel cells, is an important milestone towards sustainable energy production. To this end, heteroatom-doped graphenes have been explored extensively\cite{wang2012,zhang2015,agnoli2016} following experiments showing high ORR activity of a nitrogen-doped graphene (NG) electrocatalyst in 2010.\cite{qu2010}
Soon after the reports of high catalytic activity of NGs, boron-doped graphene (BG) emerged as another promising candidate for efficient ORR electrocatalysis.

Sheng \textit{et al.}\cite{sheng2011} measured favorable alkaline ORR activity for BG with 3.2~\% dopant concentration synthesized using Hummer's method.\cite{hummers1958,marcano2010} Their BG material catalyzed the 4\ce{e-} ORR pathway and showed good tolerance to \ce{CO} poisoning. Note that Hummer's method has become subject to criticism as it can deposit significant amounts of transition metal impurities in the material\cite{wang2013,masa2015} which cannot be removed using typical wet-chemical purification methods.\cite{ambrosi2012} In the same vein, Xu \textit{et al.}\cite{xu2014} and Jiao \textit{et al.}\cite{jiao2014} synthesized NG and BG using Hummer's method. Both groups report that NG and BG are efficient ORR catalysts, showing similarly high ORR activity in their experiments and corresponding calculations. Further experimental work is summarized in a 2016 review by Agnoli and Favaro.\cite{agnoli2016}

Computational predictions of the ORR activity of BG have overall been promising. The free energy approach using the computational hydrogen electrode (CHE)\cite{norskov2004} is typically used to evaluate the ORR activity of computational models. Since the estimate of an overpotential obtained by this approach only reflects thermodynamic free energy of intermediates as well as initial and final states, it will be referred to as the thermochemical overpotential, \tcm, in the following.

Jiao and co-workers predict a \tcm\ range of 0.4--0.6~V for both BG and NG based on calculations using the B3LYP functional and molecular flake model systems, in good agreement with their experimental measurements.\cite{jiao2014} A similar value, 0.38~V, is reported by Wang \textit{et al.} for a BG nanoribbon\cite{wang2016} using the PBE functional and DFT-D3\cite{grimme2010,grimme2011} dispersion correction. The most optimistic prediction is reported by Fazio and co-workers with a \tcm\ of 0.29~V in a B3LYP-based study of a BG flake model system.\cite{Fazio2014} For reference, the measured overpotential of a typical Pt/C electrocatalyst is 0.3--0.4~V.\cite{vielstich_handbook_2003} The experimental overpotential, however, depends on many other factors besides adsorption strength of the ORR intermediates, hence \tcm\ values are only a rough and purely thermodynamic estimate of the actual overpotential.

The exact mechanism of the ORR on BG is a matter of ongoing investigation. Fazio and co-workers established that the associative \ce{4e-} pathway should be dominant for BG from a theoretical perspective.\cite{Fazio2014} They found \ce{O2} adsorption to occur via an open-shell end-on intermediate using a molecular flake model system and the B3LYP functional. Ferrighi \textit{et al.} proposed the formation of stable \ce{B-O3} bulk oxides on BG which they hypothesize to be the first step in the ORR mechanism on BG.\cite{ferrighi2014} They, however, did not detail further reaction steps. Ferrighi \textit{et al.} used a molecular flake model and the B3LYP functional as well as periodic surface models and the PBE functional in their study. Contrarily, Wang and co-workers recently identified a cluster of two B dopants in \textit{para} arrangement to enable the associative \ce{4e-} ORR pathway, including energetically favorable \ce{O2} adsorption.\cite{wang2016} They used a periodic nanoribbon model and the PBE functional with DFT-D3 dispersion correction. Using a molecular flake model and the B3LYP functional, the study by Jiao \textit{et al.}\cite{jiao2014} finds that a top adsorption geometry should be favored for the critical *O intermediate on BG while other studies\cite{ferrighi2014,Fazio2014,wang2016} typically find a B-C bridge site to be favored for *O adsorption. It can be summarized that the active site debate for the ORR mechanism on BG is not settled yet.

Furthermore, the stabilization of the ORR intermediates on BG by water molecules, which has been found to be critical to correct energetic description of the ORR on NG,\cite{okamoto2009,Yu2011,chai2014,Reda2018} has only been considered by one group so far to the best of the authors' knowledge. Fazio \textit{et al.} used a cluster of 6 water molecules in contact with a molecular flake model representing BG to estimate the effects of solvation.\cite{Fazio2014} The group found that while the stability of the *O intermediate is barely affected by solvation, the *OH and *OOH intermediates are stabilized by -0.37 eV and -0.46 eV, respectively. The low predicted \tcm\ of 0.29~V~\textit{vs.}~SHE in this study results in part from the stabilizing effect of solvation.

In the study by Jiao \textit{et al.}\cite{jiao2014} solvation effects are estimated using implicit\cite{CramerSOLV} solvation models. However, implicit solvation models
have in some cases been shown to fail at reproducing experimental solvation energy measurements or solvation energy results from simulations using many explicit solvent molecules.\cite{eskyner2015,zhang2017,gray2017,heenen2020}

We recently presented results for the ORR on NG where it was shown that high-level DFT calculations based on hybrid functionals yield a \tcm\ estimate close to 1.0~V \textit{vs.} SHE,\cite{Kirchhoff2021a} indicating catalytic inactivity. The choice of hybrid functional was made as a result of benchmarking against a diffusion Monte Carlo data set. Generalized gradient approximation functionals were found to underestimate \tcm\ by up to 0.4 eV, thereby indicating much too high catalytic activity. However, it was noted that solvation effects could considerably improve the catalyst activity predictions. To illustrate this effect, we applied two sets of solvation stabilization energy, \DDEsolv, data for the ORR intermediates on NG taken from literature sources (Reda \textit{et al.}\cite{Reda2018} and Yu \textit{et al.}\cite{Yu2011}).
and found \tcm\ to be reduced by up to 0.5~V.
However, the published \DDEsolv\ data set were calculated in different ways and disagreed significantly, leading to different \tcm\ estimates depending on the choice of \DDEsolv\ data set.

The accurate hybrid DFT approach was also applied to BG with similar results: a \tcm\ estimate above 1.0~V \textit{vs.} SHE, indicating catalytic inactivity.\cite{Kirchhoff2021Diss} This result is in stark contrast to other more optimistic studies which, importantly, used functionals such as PBE and B3LYP as well as molecular flake models which were shown to produce unreliable adsorption free energy results.\cite{Kirchhoff2021a} However, the high \tcm\ prediction for BG did not include any solvation effects. Informed by the report from Fazio \textit{et al.} on the significant impact of \DDEsolv\ on the free energy trends and by our own observations of the same for NG, the present study was conceived to systematically investigate the effect of an increasing number of explicit water molecules on the stability of the ORR intermediates *O, *OH, and *OOH, as represented by the \DDEsolv\ descriptor.
%
Simulations were performed with the 32-atom BG model system used previously\cite{Kirchhoff2021Diss} in contact with up to 4 layers (32 molecules) of water. Both local minimization calculations as well as regular and coarse-grained classical dynamics simulations were performed using atomic forces estimated
from density functional theory (DFT) calculations to obtain statistical estimates of \DDEsolv. Additionally, local optimization calculations were performed on structures re-sampled from these data sets. In short, none of the data sets generated in this way yielded converged and trustworthy \DDEsolv\ results. Technical aspects of the simulations are discussed in detail and the conclusion is that a much larger number of water molecules needs to be included in the calculations to
provide reliable estimates of the solvation effect.
%
The present model system includes up to 139 atoms and the dynamics simulations span up to 100 ps,
thereby already straining the computational resources.
Moreover, the \DDEsolv\ estimates are highly system dependent and would need to be reestablished for every new (electro-) catalyst model.
Hence, we highlight the need for hybrid simulation methods that enable simulations of systems including hundreds or even thousands of water molecules
using a lower level of theory while retaining electronic structure level accuracy in the surface region where reactions occur.
%


\section{Methodology}\label{s:methodology}

\subsection{Calculation of the solvation stabilization energy} \label{s:ddesolv}

The solvation stabilization energy $\Delta \Delta E_{\mathrm{solv}}$ is estimated as the difference between the adsorption energy calculated for models in contact with explicit solvent ($\Delta E_{\mathrm{ads}}^{\mathrm{with\ solvent}}$) and models without inclusion of any solvent molecules ($\Delta E_{\mathrm{ads}}^{\mathrm{without\ solvent}}$):
\begin{equation}
    \Delta \Delta E_{\mathrm{solv}} = \Delta E_{\mathrm{ads}}^{\mathrm{\,with\ solvent}} - \Delta E_{\mathrm{ads}}^{\mathrm{\,without\ solvent}}, \label{eq:ddesolv_long}
\end{equation}
where
\begin{eqnarray}
    \Delta E_{\mathrm{ads}}^{\mathrm{\,with\ solvent}} &=& E_{\mathrm{tot}}^{\mathrm{\,BG\ +\ adatom \ with\ solvent}} - E_{\mathrm{tot}}^{\mathrm{\,BG\ with\ solvent}} \nonumber \\
    & & - E_{\mathrm{tot}}^{\mathrm{\,adatom\ reference}}
\end{eqnarray}
and
\begin{eqnarray}
    \Delta E_{\mathrm{ads}}^{\mathrm{\,without\ solvent}} &=& E_{\mathrm{tot}}^{\mathrm{\,BG\ +\ adatom \ without\ solvent}} - E_{\mathrm{tot}}^{\mathrm{\,BG\ without\ solvent}} \nonumber \\
    & & - E_{\mathrm{tot}}^{\mathrm{\,adatom\ reference}}.
\end{eqnarray}

Here, $E_{\mathrm{tot}}^{\mathrm{adatom\ reference}}$ is the total energy of any combination of gasphase molecules used to calculate the adsorption energy. For example, $E_{\mathrm{tot}}^{\mathrm{adatom\ reference}}$ may be expanded to $E_{\mathrm{tot}}^{\mathrm{H2O}} - E_{\mathrm{tot}}^{\mathrm{H2}}$ to serve as the reference energy for an \*O adatom. Because these values are always gasphase reference energy values, also in the case of the solvated model systems, they cancel out in the $\Delta \Delta E_{\mathrm{solv}}$ calculation.

Therefore, equation \eqref{eq:ddesolv_long} reduces to:
\begin{eqnarray}
    \Delta \Delta E_{\mathrm{solv}} &=& E_{\mathrm{tot}}^{\mathrm{\,BG\ +\ adatom \ with\ solvent}} - E_{\mathrm{tot}}^{\mathrm{\,BG\ with\ solvent}} \nonumber \\
    & & - (E_{\mathrm{tot}}^{\mathrm{\,BG\ +\ adatom \ without\ solvent}} - E_{\mathrm{tot}}^{\mathrm{\,BG\ without\ solvent}}) \label{eq:ddesolv_short}
\end{eqnarray}

\subsection{Calculation of the confidence interval for average ensemble properties}

The confidence interval (CI) is a useful statistical measure for the error bar of an average result sampled from a normal distribution of values. It is therefore also useful to estimate the error bar of ensemble averages sampled through molecular dynamics integration; see Grossfield \textit{et al.}\cite{Grossfield2019} for more details. The CI defines an interval in which the true ensemble average lies with a certain probability. Here, a 95 \% probability threshold is used to define the error bars, \textit{i.e.}, the 95 \% CI.

The two-sided CI $<x>$ of a variable $x$ is defined as
\begin{equation}
    <x> = \bar{x} \pm U, \label{eq:ci1}
\end{equation}
where $\bar{x}$ is the ensemble average and $U$ is the expanded uncertainty.
The expanded uncertainty is defined as
\begin{equation}
    U = k\ s(\bar{x}),
\end{equation}
where $k$ is the coverage factor and $s(\bar{x})$ is the experimental standard deviation of the mean.
$s(\bar{x})$ is defined as
\begin{equation}
    s(\bar{x}) = \frac{s(x)}{\sqrt{n}}, \label{eq:ci3}
\end{equation}
where $s(x)$ is the experimental standard deviation
\begin{equation}
    s(x) = \sqrt{\frac{\sum_{j=1}^{n} (x_j - \bar{x})^2}{n -1}} \label{eq:ci4}
\end{equation}
with the sample values $x_j$, the arithmetric mean of the ensemble property $\bar{x}$, and the number of independent samples $n$.

The coverage factor $k$ is a measure for the number of independent samples taken into account during calculation of the standard deviation. For the 95 \% CI used in this work, the coverage factors $k$ are given by Grossfield \textit{et al.} as follows:

\begin{table}[htbp]
    \centering
    \renewcommand{\arraystretch}{1.5}
    \caption{Coverage factors $k$ as a function of the number of independent samples $n$. Reproduced from Grossfield \textit{et al.}\cite{Grossfield2019}}
    \begin{tabular}{|l|l|}
    \hline
        $n$ & $k$ \\ \hline
        6 & 2.57 \\
        11 & 2.23 \\
        16 & 2.13 \\
        21 & 2.09 \\
        26 & 2.06 \\
        51 & 2.01 \\
        101 & 1.98 \\ \hline
    \end{tabular}
    \label{t:coverage_factors}
\end{table}

\section{Computational Details}\label{s:computational}

\subsection{BG sheet model system}

\begin{figure}[htbp]
    \centering
    \includegraphics[width=0.5\linewidth]{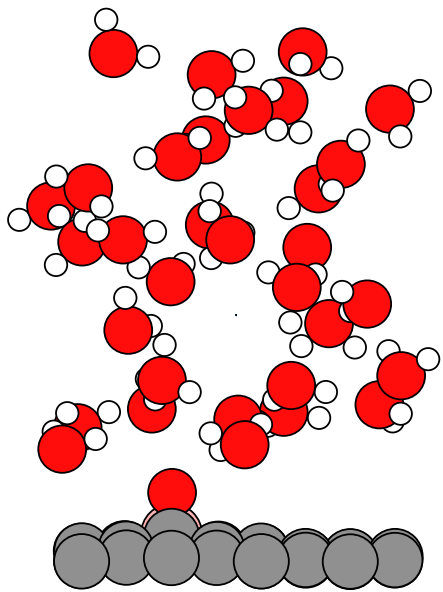}
    \caption{Rendered illustration of the BG sheet model system with an *O adatom in contact with 32 water molecules (4 layers).}
    \label{f:bg_model_solv}
\end{figure}

The model system used in this study is a 32-atomic graphene sheet with one B dopant atom, analogous to our previous works on NG and BG.\cite{Kirchhoff2021a, Kirchhoff2021Diss} To study the influence of solvation on the ORR intermediates *O, *OH, and *OOH, 1-4 layers of water molecules with 8 water molecules per layer are added to the model. The water configurations built initially were inspired by the configurations presented by Reda \textit{et al.} in a study of the solvation of ORR intermediates on NG.\cite{Reda2018} The group showed that the maximum H$_2$O coverage per layer for NG is $\Theta_ \mathrm{H_2O} = \frac{2}{3}$ monolayers which the present results confirm. Hence, a maximum of 24 atoms (8 molecules) can be placed per layer before lateral crowding destabilizes the water configuration and formation of a new layer begins. Figure \ref{f:bg_model_solv} shows a representative illustration of the BG sheet model with an *O adatom in contact with 4 layers of water molecules; illustrations of sheet models with *OH and *OOH admolecules as well as models in contact with 1-3 layers of water are shown in figures S1 and S2, respectively.

In agreement with studies by Fazio \textit{et al.}, \cite{Fazio2014} Ferrighi \textit{et al.},\cite{ferrighi2014} and Wang \textit{et al.}\cite{wang2016} but in disagreement with the study by Jiao \textit{et al.},\cite{jiao2014} we find adsorption of the *O intermediate on the C-B bridge position to be energetically most favorable. The *OH and *OOH adspecies are found to adsorb most favorably on the B top position, which is in agreement with all previously mentioned studies.

The 32-atomic BG model system is converged with respect to the adsorption energy of the ORR intermediates, *O, *OH, and *OOH, see figure S4. This model therefore allows for the study of the adsorption energy - and the influence of solvation thereon - for a dilute system where the electronic effects of both the dopant atom and the adspecies are isolated and crowding effects can be ruled out.

\subsection{Simulation parameters}

The obtained data sets, including input files with simulation parameters, are distributed alongside this article and are available under DOI:10.5281/zenodo.7684918.

\subsection{Choice of DFT code and functional}

All simulations were performed with the VASP software version 6.2.0.\cite{kresse1993, kresse1994, kresse1996, kresse1996_2} All calculations used the RPBE density functional\cite{hammer1999} with DFT-D3 dispersion correction.\cite{grimme2010,grimme2011} The RPBE-D3 method has been shown to yield water configurations in good agreement with experiments and higher-level methods at comparatively low computational cost.\cite{tonigold2014} Previous work on NG showed that adsorption energy values for the ORR intermediates can be wrong by up to 0.4~eV compared to the best estimate provided by the HSE06 hybrid functional, which was found to give the lowest error of 5~\% compared to a diffusion Monte Carlo benchmark calculation.\cite{Kirchhoff2021a} Similar results were obtained for BG,\cite{Kirchhoff2021Diss} see table S1, where \tcm\ with the HSE06 functional was \textit{ca.}~1.0~V \textit{vs.} SHE and GGA functionals underestimated this best-estimate value by up to 0.6 V. Figure S3 shows the free energy trends for the ORR on BG obtained with various density functionals. However, our previous work also showed that \DDEsolv\ does not share the same strong dependency on the functional.\cite{Kirchhoff2021a} This realization enables the present study since FPMD simulations as long as required for this work are currently not computationally feasible using hybrid functionals.

\subsubsection{Static DFT calculations} \label{s:static}

Static calculations constitute single-point electronic energy calculations as well as minimization of the total energy with respect to the atomic coordinates. Wave functions were self-consistently optimized until the energy in subsequent iterations changed by less than $10^{-6}$~eV. The wave function was sampled using Monkhorst-Pack \textit{k} point grids.\cite{monkhorst1976} A \textit{k} point density larger than 2$\times$2$\times$1 was found to give converged results for $\Delta \Delta E_\mathrm{solv}$, see figure S5. Due to the wide variety of structures calculated in this work, refer to the data set distributed alongside this article to see the chosen \textit{k} point density for each subset of calculations.

Simulations were carried out using a plane wave basis set with an energy cutoff of 600~eV to represent valence electrons and the projector-augmented wave (PAW) method\cite{blochl_projector_1994,kresse1999} was used to account for the effect of inner electrons. See figure S6 for a convergence study for the PAW energy cutoff. Gaussian-type finite temperature smearing was used to speed up convergence. The smearing width is chosen so that the electronic entropy was smaller than 1~meV in all cases. Real-space evaluation of the projection operators was used to speed up calculations of larger systems, using a precision of $10^{-3}$~eV~atom$^{-1}$. Atomic coordinates were optimized until forces reached below $10^{-2}$~eV~\AA$^{-1}$. The L-BFGS limited-memory Broyden optimizer from the VASP Transition State Tools (VTST) software package was used to minimize the forces with respect to the atomic coordinates. The periodic images are separated by 14~\AA\ of vacuum and a dipole correction is applied perpendicular to the slab.

\subsubsection{Classical molecular dynamics simulations}

Classical molecular dynamics (MD) simulations were carried out in an \textit{NVT} ensemble at 300~K using the Langevin dynamics\cite{vanden2006} implemented in VASP.
The simulations used similar parameters to those outlined in section \ref{s:static} but used a lower PAW energy cutoff of 400~eV and a 3$\times$3$\times$1 Monkhorst-Pack \textit{k} point grid for computational efficiency. A Langevin friction parameter of $\gamma = 4.91$ was used throughout all simulations.

Dynamics were calculated initially until the total energy and temperature were converged. This equilibration period is not considered in the evaluation and was optimized on a case-by-case basis. After equilibration had been achieved, the actual sampling was performed over a period of time. In all simulations the geometry of the graphene sheet and the adspecies were constrained to the geometry obtained from a one-shot geometry optimization of the system in contact with $n = 1-4$ water layers, respectively. Only the water molecules were allowed to move during simulations. The $E\mathrm{tot}$ \textit{vs.} $t$ and $T$ \textit{vs.} $t$ trends for all simulations are shown in the online SI.

Two data sets were generated:

\begin{enumerate}
    \item First, simulations were performed without any constraints on the water molecules and with a time step of 0.1 fs. Simulations were continued up to a total simulation time of 10 ps after thermalization. This set of MD simulations will be referred to as the \textit{flexible MD} data set going forward.

    \item Second, simulations were repeated after placing a Rattle-type bond length constraint\cite{Andersen1983} on the O-H and H-H bonds to keep the geometry of water molecules rigid throughout simulations, thus enabling a coarse-grained time step of 1.0 fs. Simulations were continued up to a total simulation time of 100 ps after thermalization. This set of MD simulations set will be referred to as the \textit{constrained MD} data set going forward.
\end{enumerate}

To obtain \DDEsolv, configurations were sampled every 1~ps, yielding 10 samples for the \textit{flexible MD} data set and 100 samples for the \textit{constrained MD} data set. This choice of sampling frequency is informed by the correlation time of water. The correlation time is the time it takes for complete re-orientation of the water arrangement, thus yielding a new, independent sample configuration that is statistically significant. It was found to be \textit{ca.}~1.7~ps for water at room temperature using nuclear magnetic resonance spectroscopy.\cite{Lankhorst1982} The chosen sampling rate of 1~ps is smaller than this value as a result of the significant computational effort of performing long dynamics simulations. To minimize the risk of oversampling, Langevin dynamics was chosen to describe coupling to a heat bath. Langevin dynamics introduces a stochastic component to the propagation which can help to diversify configurations more quickly compared to fully deterministic dynamics.

\section{Results}\label{s:results}

\subsection{One-shot minimization of atomic coordinates} \label{s:oneshot}

\begin{figure}[htbp]
    \centering
    \includegraphics[width=\linewidth]{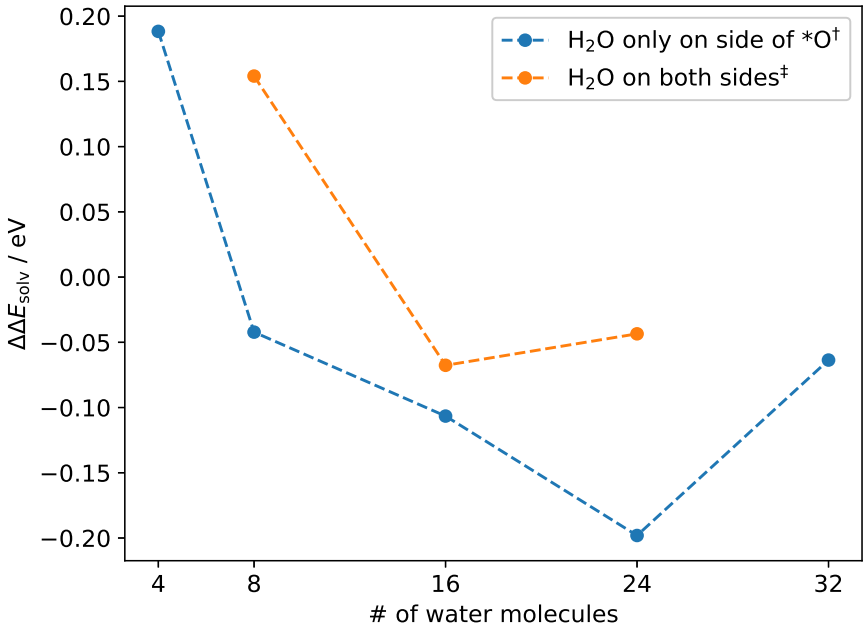}
    \caption{\DDEsolv\ results for the *O intermediate on BG in contact with 4--32 molecules of water obtained from the \textit{one-shot minimization} data set. The blue line shows \DDEsolv\ when water molecules are exclusively placed on the side of the model where the adatom is located. The orange line shows \DDEsolv\ values from select models where water molecules are placed on both sides of the model. For the orange line, the $x$ axis indicates the number of water molecules on the side with the adatom and not the total number of water molecules. The $\dagger$ and $\ddagger$ indicators connect the values in this figure to the corresponding data values in table \ref{t:ddesolv_oneshot}.}
    \label{f:ddesolv_oneshot}
\end{figure}

The first data set is generated by bringing the BG model system with *O, *OH, and *OOH adspecies into contact with 4--32 molecules of water and minimizing the resulting configurations with respect to the atomic forces. This data set will be referred to as the \textit{one-shot minimization} data set going forward. The chosen water configurations are modeled after those used by Reda \textit{et al.} to calculate the solvation stabilization energy for the ORR intermediates on NG sheet model systems.\cite{Reda2018} Configurations were created so that water molecules are only on one side of the BG sheet model or on both sides, denoted with the $\dagger$ and $\ddagger$ symbols, respectively, in table \ref{t:ddesolv_oneshot} and figure \ref{f:ddesolv_oneshot}.
\begin{table}[htbp]
    \caption{Summary of the calculated $\Delta \Delta E_\mathrm{solv}$ results based on the \textit{one-shot minimization} data set. $^{\dagger}$ and $^{\ddagger}$ indicate values where water molecules are place only on one side or on both sides of the BG sheet model, respectively.}
    \centering
    \renewcommand{\arraystretch}{1.5}
    \begin{tabular}{|l|l|l|}
    \hline
        \# of water molecules & Arrangement & $\Delta \Delta E_\mathrm{solv}$ / eV \\ \hline
        4 H$_2$O & on side with *O & 0.19$^{\dagger}$ \\
        8 H$_2$O & on side with *O & -0.04$^{\dagger}$ \\
        16 H$_2$O & on side with *O & -0.11$^{\dagger}$ \\
        16 H$_2$O & 8 on both sides & 0.15$^{\ddagger}$ \\
        24 H$_2$O & 16 on side with *O, 8 on the other side & -0.07$^{\ddagger}$ \\
        24 H$_2$O & on side with *O & -0.20$^{\dagger}$ \\
        32 H$_2$O & 24 on side with *O, 8 on the other side & -0.04$^{\ddagger}$ \\
        32 H$_2$O & on side with *O & -0.06$^{\dagger}$ \\
        32 H$_2$O & 16 on side with *O, 16 on the other side & -0.06 \\ \hline
    \end{tabular}
    \label{t:ddesolv_oneshot}
\end{table}

The \DDEsolv\ results obtained from the \textit{one-shot minimization} data set give rise to several trends. First, when water molecules are placed only on one side of the model, \DDEsolv\ for the *O intermediate does not appear to be converged within the tested series of models as \DDEsolv\ still increases from -0.20 eV to -0.06 from 24 to 32 molecules. Values can be deemed converged if changes are below \textit{ca.} 0.05 eV or 1 kcal~mol$^{-1}$, \textit{i.e.}, chemical accuracy.

Second, the results for simulations where molecules are placed only on the side of the sheet model with the adatom ($\dagger$) are inconsistent with simulations where molecules are placed on both sides of the model ($\ddagger$). For example, deviations of < 0.05 eV are found between simulations where 16 molecules are placed on the side of the adatom and 0, 8, and 16 molecules are placed on the other side. This result would potentially indicate that water molecules on the opposite side of where the adspecies is located have negligible influence and can be omitted. However, the deviation between \DDEsolv\ values where 8 molecules are placed on the side with the adspecies and 0 or 8 molecules are placed on the other side is 0.19 eV. Similarly, the deviation between \DDEsolv\ values where 24 molecules are placed on the side with *O and 0 or 8 molecules are placed on the other side is 0.16 eV.

Results from \textit{one-shot minimization} data set are therefore inconsistent. From this data, it is unclear if and when \DDEsolv\ will converge as a function of the number of added water molecules and it cannot be assessed with confidence if water molecules do or do not need to be present on the side of the sheet opposite of the adspecies.

One potential reason for the inconsistent behavior lies in the one-shot nature of the data set: water molecule arrangements are flexible and form a complex energy landscape where minimization algorithms can easily become stuck in local minimum configurations. This limitation can be overcome by rigorous sampling of the configurational space by MD integration.

\subsection{\textit{NVT} simulations}

\begin{figure}[htbp]
    \centering
    \includegraphics[width=\linewidth]{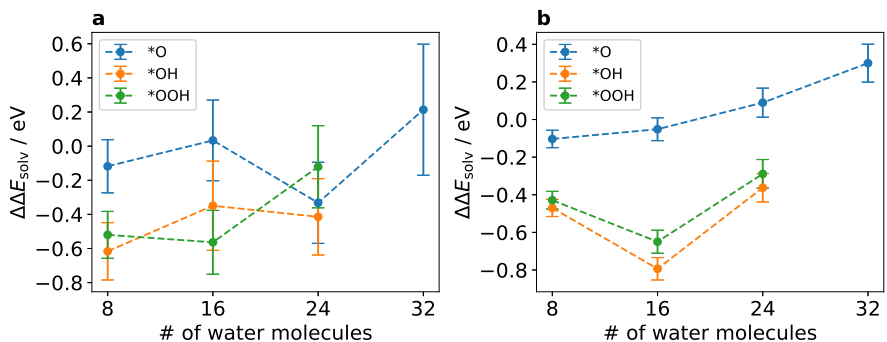}
    \caption{\DDEsolv\ results for the *O (blue curve), *OH (orange curve), and *OOH (green curve) adspecies on BG in contact with 8--32 molecules of water obtained as ensemble averages from \textbf{a} 10 ps of MD using a time step of 0.1 fs where water molecules were flexible and \textbf{b} 100 ps of MD using a time step of 1.0 fs where water molecules were constrained. The error bars indicate the two-sided 95 \% CI calculated according to equations \eqref{eq:ci1}-\eqref{eq:ci4}.}
    \label{f:ddesolv_md}
\end{figure}

In order to probe if insufficient sampling of the configurational space is responsible for the inconsistent results of the \textit{one-shot minimization} data set, \DDEsolv\ is subsequently determined as an ensemble average by performing MD simulations for a total of 10 ps using a time step of 0.1 fs. No constraint was placed on the O-H and H-H bonds of water molecules. This set of simulations is referred to as the \textit{flexible MD} data set. Due to the significant computational effort of these simulations, only water configurations where water molecules are placed on the side of the adspecies are considered. Simulations are performed for the clean BG sheet model, for the BG sheet with an *O adatom in contact with 8--32 molecules, and for the *OH and *OOH adspecies in contact with 8--24 molecules of water. Figure \ref{f:ddesolv_md}\textbf{a} visualizes the \DDEsolv\ results calculated from this data set.

Focusing on the *O intermediate (blue curve), a similar trend of \DDEsolv\ \textit{vs.} the number of water molecules emerges as before from the \textit{one-shot minimization} data set: the values oscillate and there is an increase of \DDEsolv\ from -0.3 eV to 0.2 eV from 24 to 32 molecules, indicating significant destabilization of this adspecies. In general, the differences between subsequent data points are found to be larger than in the case of the \textit{one-shot minimization} data set.

It can be summarized that the \textit{flexible MD} data set did not yield more consistent \DDEsolv\ results than the \textit{one-shot minimization} data set. While a similar overall \DDEsolv\ trend is observed for the *O adspecies, differences between subsequent data points are larger than in the case of the \textit{one-shot minimization} data set.

Another important observation is the significant sizes of the error bars, which extend from 0.25 eV up to over 0.5 eV in some cases. Note that in the case of the *O intermediate, the error bar span becomes larger as a function of the number of water molecules. This effect is much less pronounced, if at all, for the *OH and *OOH intermediates. However, it is clear from the size of the error bars that the length of simulation time is too short compared to the correlation time of water and thus simulations only yielded 10 independent samples that entered into the evaluation.

In an effort to extend the simulation time, a coarse-graining approach was chosen where the O-H and H-H bond lengths of water molecules were constrained to the average corresponding bond lengths obtained in the \textit{flexible MD} data set. This bond length constraint allows for larger simulation time steps to be taken without the risk of spurious discretization errors from inadequate sampling of the fast O-H vibrations. A subsequent set of dynamics simulations of the same model systems thus used a time step of 1.0 fs and was continued for a total of 100 ps simulation time, yielding 100 independent samples. \DDEsolv\ results from this \textit{constrained MD} data set are visualized in figure \ref{f:ddesolv_md}\textbf{b}.

\DDEsolv\ trends from the \textit{constrained MD} data set, while also showing no signs of converging behavior, differ significantly from the \textit{flexible MD} and \textit{one-shot optimization} data sets. The obtained \DDEsolv\ values for the *O adspecies do not oscillate as in the case of the other data sets but continuously increase with increasing number of water molecules. From this data set, the presence of 24 and 32 water molecules is predicted to significantly destabilize this intermediate. With \textit{ca.} 0.25 eV, the data point for 32 water molecules from this data set is similar to the \textit{flexible MD} data set, however, this data set does not show the reduction of \DDEsolv\ at 24 molecules that was observed for both the \textit{flexible MD} and the \textit{one-shot minimization} data sets.

The *OH and *OOH adspecies show similar \DDEsolv\ trends that parallel each other in this data set; however, values oscillate by up to 0.5 eV when the number of water molecules is increased. Finally, the factor 10 longer simulation time affects the size of the error bars which is now on the scale of \textit{ca.} 0.1 eV. Similar to results from the \textit{flexible MD} data set, the error bars for \DDEsolv\ of the *O adspecies are found to increase with increasing number of water molecules in the simulation while no such trend is observed for the *OH and *OOH intermediates.

Finally, the local structure of the water molecules around the adspecies is analyzed using $z$ distribution function, $g(z)$, see figure S7. The $g(z)$ distributions are obtained by calculating distances between the O atoms of water molecules and an $x-y$ plane within the BG sheet model. The $g(z)$ show distinct bands for the first and second solvation layer. The bands for 3 and 4 layers are significantly more broadened, indicating that the surface-adjacent double layer is more strongly coordinated compared to subsequent layers. Notably, shoulders at the first band are visible in the $g(z)$ from the \textit{flexible MD} data set which are not visible in the \textit{constrained MD} data set. However, this result is presented with the caveat that the data is more noisy compared to the smoother \textit{constrained molecule} $g(z)$ results due to the 10x smaller sampling statistics. This result potentially indicates that the bond length constraint affects the coordination fine structure around the adspecies and thus may help to explain the differences between the \textit{flexible MD} and \textit{constrained MD} data sets. However, more detailed investigation is required to validate the importance of this observed difference.

It can be summarized that coarse-grained MD simulations yielded a data set that is significantly different from the more similar-to-each-other \textit{flexible MD} and \textit{one-shot minimization} data sets but did not yield more consistent \DDEsolv\ results either. Finally, the bond length constraint is found to change the \DDEsolv\ results compared to the \textit{flexible MD} data set; however, since there are currently no converged reference values for \DDEsolv, it is impossible to assess if the changes introduced by the Rattle-type constraint are detrimental or not.

\subsection{Re-sampling and energy minimization}

\begin{figure}[htbp]
    \centering
    \includegraphics[width=\linewidth]{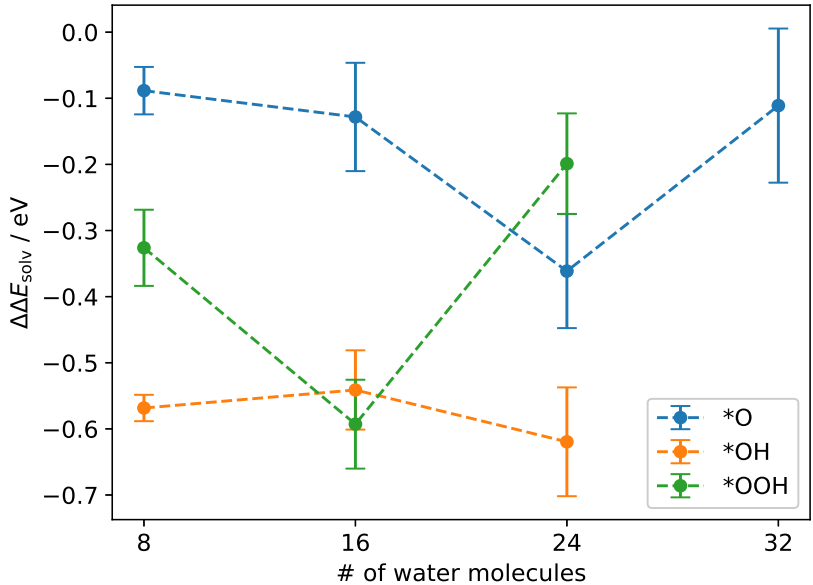}
    \caption{\DDEsolv\ results for the *O, *OH, and *OOH adspecies on BG in contact with 8--32 molecules of water obtained as average values over 20 images per data point which were randomly resampled from the flexible MD data set and subsequently energy-minimized with respect to the atomic coordinates. The error bars indicate the two-sided 95 \% CI calculated according to equations \eqref{eq:ci1}-\eqref{eq:ci4}.}
    \label{f:ddesolv_resampled}
\end{figure}

The \textit{flexible MD} and \textit{constrained MD} data sets did not yield converged \DDEsolv\ results. There are, however, two technical limitations which may reduce the significance of these data sets:
\begin{enumerate}
    \item For these data sets, \DDEsolv\ is calculated by using the average total energy from an \textit{NVT} ensemble ($T = 300$~K) for the energy terms labeled "with solvent" in equation \eqref{eq:ddesolv_short}. The energy terms labeled "without solvent" are obtained from energy minimization calculations of the systems without solvent which are technically at 0~K temperature. While the BG sheet model and adspecies were kept frozen in the atomic configuration from a 0~K energy minimization during the MD and only water molecules were allowed to move, it cannot be fully excluded that results are biased due to a mismatch between the averaged finite-temperature MD values on one side and the locally optimized, 0~K values on the other side of the equation.
    \item As outlined in section \ref{s:computational}, the MD simulations - as well as the corresponding reference simulations of the systems "without solvent" needed for equation \eqref{eq:ddesolv_short} - used a reduced PAW energy cutoff value of 400~eV to enable longer simulation times. This value is technically not converged for adsorption energy calculations, see figure S5.
\end{enumerate}
In order to address both of these limitations, a fourth data set is produced. To this end, 20 structures are randomly sampled from each \textit{flexible MD} trajectory and subsequently energy-minimized using the settings presented in section \ref{s:computational}, \textit{i.e.}, with a larger PAW energy cutoff of 600 eV. This way, the diversity of the MD-generated configurations is maintained but all values entering equation \eqref{eq:ddesolv_short} are obtained from energy-minimized atomic configurations using safe accuracy settings. This data set will be referred to as the \textit{resampled} data set going forward. Figure \ref{f:ddesolv_resampled} visualizes the \DDEsolv\ results obtained from this data set.

The \textit{resampled} data set shares similarities with the \textit{flexible MD} and \textit{one-shot optimization} data sets, for example the characteristic dip of \DDEsolv\ for the *O adatom at 24 water molecules. This result further indicates that the bond length constraint used to obtain the \textit{constrained MD} data set is likely altering the trends in a significant way. The previously discussed trend regarding error bar spans increasing with increasing number of molecules is distinctly present both for the *O and the *OH adspecies. Ultimately, this data set does not provide fundamentally different insights into the \DDEsolv\ trends compared to the preceding analyses.

\section{Discussion}\label{s:discussion}

\subsection{Comparison of the
results
from different data sets}

\begin{figure}[htbp]
    \centering
    \includegraphics[width=\linewidth]{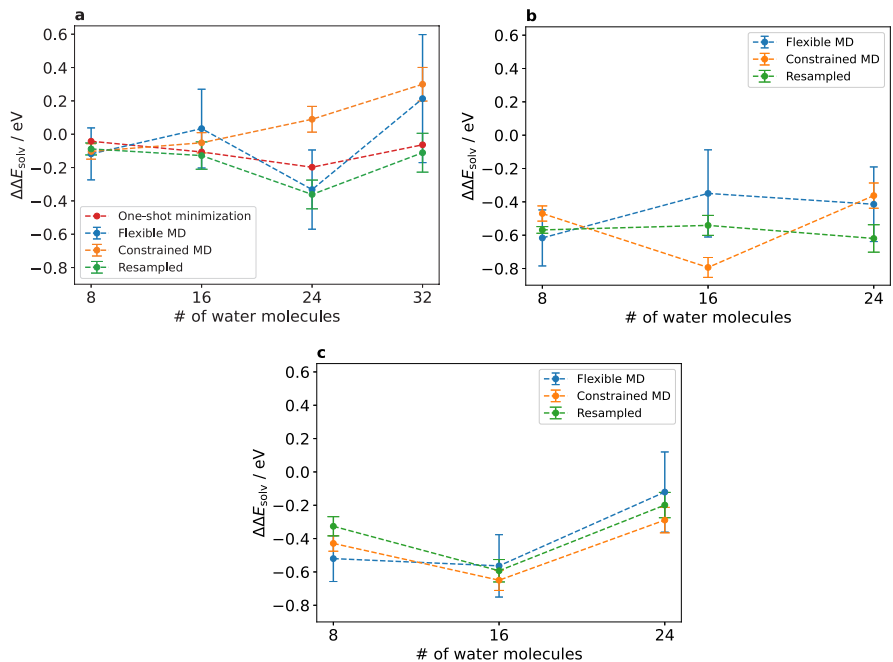}
    \caption{Comparison of $\Delta \Delta E_\mathrm{solv}$ results for the \textbf{a} *O, \textbf{b} *OH, and \textbf{c} *OOH adspecies from the \textit{one-shot minimization} data set, the \textit{flexible MD} and \textit{constrained MD} data sets, and the \textit{resampled} data set.}
    \label{f:ddesolv_comparison}
\end{figure}

Figure \ref{f:ddesolv_comparison} shows a side-by-side comparison of \DDEsolv\ as a function of the number of water molecules for the *O, *OH, and *OOH adspecies from the four obtained data sets.
The \textit{resampled} data set is the most significant data set among those obtained in this work as it combines the broad configurational diversification of the MD simulations with the methodological consistency of calculating \DDEsolv\ using strict accuracy parameters and exclusively on the basis of energy-minimized structures. By comparing the data sets with each other and with the \textit{resampled} data set in particular, several important aspects can be highlighted.

First, convergence of \DDEsolv, \textit{i.e.} changes of < 0.05 eV between subsequent data points, is not observed in any case. It is impossible at this point to give a confident estimate of \DDEsolv\ for the tested adspecies on the BG sheet model. This result indicates that more than 32 molecules (4 layers) of water are likely necessary to obtain converged results.

Converging the \DDEsolv\ value to changes within chemical accuracy is of crucial importance. For example, consider the potential-dependent free energy trends for the ORR on the BG model presented in figure S3. These trends were obtained according to the free energy approach using the computational hydrogen electrode.\cite{norskov2004} Using the most reliable functional for adsorption energy calculations on this material class according to benchmarks,\cite{hsing2012,janesko2013,Kirchhoff2021a} the HSE06 hybrid functional, the potential-determining step is the formation of the *OOH intermediate by a significant margin. The extrapolated thermochemical overpotential, \tcm, for the ORR on the present BG model is \textit{ca.}~1.0V~\textit{vs.}~SHE. Stabilization of the *OOH intermediate by roughly -0.4 eV (8 water molecules), -0.6 eV (16 water molecules), or -0.2 eV (24 water molecules) will therefore proportionally reduce \tcm\ to 0.6~V, 0.4~V, and 0.8~V~\textit{vs.}~SHE, respectively. Depending on the number of included water molecules, one can predict a mostly inactive (\tcm\ = 0.8~V, 24 molecules) or moderately active (\tcm\ = 0.4~V, 16 molecules) ORR electrocatalyst. The overpotential of a typical reference Pt/C electrocatalyst is 0.3--0.4~V.\cite{vielstich_handbook_2003} Therefore, \DDEsolv\ must be converged within the limits of chemical accuracy before any trustworthy prediction can be made.

Second, there appears to be no obvious systematicity to whether trends from the different data sets agree with each other or not. For example, values from different data sets for the *OOH intermediate are in reasonable agreement and show similar overall trends. In the case of the *O adatom, there is some correlation between trends from correlated data sets (in particular the \textit{flexible MD} data set and the \textit{resampled} data set which was generated from the former) and only the \textit{constrained MD} data set behaves significantly different. In the case of *OH, however, there appears to be no shared trends between results from either of the data sets. Further research is needed to analyze why there is reasonable agreement in some cases and no agreement in other cases.

Third, the error bars in all cases are significantly larger than chemical accuracy ($\pm$~0.05~eV). Aside from the fluctuation amplitude of the total energy values, the size of the error bar is governed by the number of independent samples. Because of the long experimentally measured correlation time of water, significantly longer statistics may be required to reduce the uncertainty to within chemical accuracy. See also section \ref{s:sampling} for a detailed analysis of the influence of sampling frequency.

Fourth, from the results presented in table \ref{t:ddesolv_oneshot}, it cannot be completely ruled out that water molecules may have to be added to both sides of the BG sheet model to obtain correct results. This result stands in contrast to results by Reda \textit{et al.} for NG where results for placing water molecules on one side or both sides of the model were close to identical.\cite{Reda2018} This result therefore shows that \DDEsolv\ values obtained for one material cannot be transferred to others, even if they are as closely related as NG and BG.

Fifth, analysis of the $z$ distributions, $g(z)$, of oxygen atoms from the water molecules based on the MD data sets provided some first evidence that the bond length constraint used to obtain the \textit{constrained MD} data set may have affected the coordination fine structure around the adspecies. However, due to the poor statistics resulting from the small required time step of the \textit{flexible MD} data set, it would be necessary to extend these simulations by a factor 5-10 to obtain enough independent samples to make sure that this observation is significant.

To the best of our knowledge, there is only one other study in literature where \DDEsolv\ values from explicit solvation were calculated for the ORR intermediates on BG. Fazio \textit{et al.} used a molecular BG flake model in contact with a cluster of 6 water molecules to obtain \DDEsolv.\cite{Fazio2014} The group used the B3LYP hybrid functional in combination with DFT-D3 dispersion correction. From this model, they obtained \DDEsolv\ values of -0.06 eV, -0.37 eV, and -0.46 eV for the *O, *OH, and *OOH intermediates. The values for *O and *OOH are in reasonable agreement with the results for 8 water molecules in the present study, which is the closest point of reference. The value for *OH is 0.15 to 0.20 eV more positive than in the present work. However, because the \DDEsolv\ values in the present work are not converged even when 32 water molecules are included, an in-depth discussion about potential reasons for the (dis-)agreement of the present results and the results by Fazio \textit{et al.} is not appropriate.

As an intermediary conclusion, the most likely explanation for the non-conversion of the \DDEsolv\ results in general, as well as for the non-systematic differences between data sets more specifically, is that significantly more water molecules need to be included in simulations. It is unclear at this point how many water molecules would be required to achieve convergence. Sakong \textit{et al.} found that 6 layers of water are needed to obtain bulk water behavior and converged work function estimates in the case of FPMD simulations of a Pt(111) surface in contact with water.\cite{Sakong2016} However, Pt(111) is a strongly-coordinating surface compared to the hydrophobic BG sheet model in the present study. Furthermore, the group tested for convergence of the work function and not for \DDEsolv\ of reaction intermediates. Hence, it is unlikely that the number of 6 necessary water layers will also be the correct number of layers to include for the present system.

For these reasons, it is currently not possible to foresee the ultimately required number of water molecules required to obtain converged \DDEsolv\ results for this system. Attempting to find this number systematically by dynamics simulations with DFT atomic forces quickly becomes computationally unfeasible; simulations for the models in contact with 32 water molecules in this work already required several weeks of computational time. Even if these considerable time and energy resources would be spent to identify this number for the present problem, such a study would have to be repeated for every new material under investigation. Even though the influence of solvation has been shown to significantly affect free energy trends, the authors are therefore convinced that such simulations cannot (yet) be performed routinely.

We have thus come to the decision to publish the present results as-is and to not continue running simulations with model systems that include more and more water molecules at ever increasing computational cost. Instead, we are currently focusing research efforts into development of a 2D periodic polarizable-embedding QMMM method that will allow for simulations with thousands of water molecules while retaining electronic structure level accuracy for the surface model and the closest few layers of water molecules. This method will use the Single Center Multipole Expansion (SCME) ansatz to describe polarization of water molecules which is crucial to accurately describe interface processes such as charge transfer.\cite{Jonsson2019, Dohn2019} Because the boundary plane between the QM and MM regions has exclusively water molecules on both sides, and because it is not necessary to describe diffusion to or from the surface to obtain \DDEsolv\ results, an efficient restrictive boundary method can be used. The SAFIRES method recently developed in our groups was build to support 2D periodic boundary conditions.\cite{Kirchhoff2021}

A publication on the technical implementation of the 2D periodic polarizable-embedding QMMM ansatz for the open-source GPAW and ASE programs is currently in preparation in our groups. The goal is to use this method to revisit the BG model system in the present work.

\subsection{Analysis of potential error sources}

To conclude the discussion of the data sets presented in this work, the following sections will rule out various potential error sources that readers familiar with dynamics simulations and the pitfalls of solvation energy calculations may be concerned about.

\subsubsection{Influence of the sampling frequency on the results} \label{s:sampling}

Configurations were sampled from the dynamics simulations at an interval of 1 ps.
It is important to ask how the \DDEsolv\ results are affected by changes of the sampling frequency. Figure S8 compares \DDEsolv\ results from the \textit{flexible MD} and \textit{constrained MD} data sets analyzed every 2 ps, 1 ps, 100 fs, and 10 fs.

The \DDEsolv\ results appear to be robust against the choice of sampling frequency. The only significant differences are observed when between sampling the \textit{flexible MD} data set every 2 ps (5 total samples) or every 1 ps (10 total samples) and faster. This difference can be attributed to the poor statistics in the case of the 2 ps sampling frequency.

The size of the error bars is affected significantly by the sampling frequency because the square root of the number of samples, $\sqrt{n}$, enters the divisor of equation \eqref{eq:ci3}. This test therefore highlights the importance of choosing a reasonable sampling frequency based on the physical properties of the system to obtain a meaningful error bar. It is easy to get lured into a false sense of security by oversampling the results to obtain small error bars.

\subsubsection{Spurious dipole and quadrupole corrections
} \label{s:dipole}

Total energy calculations were performed using dipole and quadrupole correction perpendicular to the surface to avoid interactions between periodic repetitions of the simulation box. It is known that first-row semiconductors with defects, of which BG is an example, can lead to large dipole and quadrupole moments, thus making the correction necessary. However, our simulations showed that the correction can sometimes give erroneously large corrections of several eV for unknown reasons. After re-optimizing the wave function in a single-point calculation, the correction is then found to be of a reasonable magnitude again, usually on the order of some meV.

Because it is impossible to perform this manual correction for all calculations in this work, the consistency of the results is representatively examined by analyzing the average dipole and quadrupole correction energy (and uncertainty thereof) of the \textit{resampled} data set. Figure S9 shows the results of this analysis. The average correction energy is <= 0.02 eV in all cases, which better than chemical accuracy. Error bars are found to be as large as 0.01 eV in some cases and close to 0.02 eV in one extreme case (BG-OOH in contact with 24 water molecules), indicating that the dipole and quadrupole energy correction is indeed volatile (in relation to the absolute values) and dependent on the exact geometry of the system. However, due to the small overall magnitude of the correction, it can be concluded that this correction should not significantly influence the calculation results.

\subsubsection{Spurious dispersion correction
}

DFT-D3 dispersion correction values are significantly larger in magnitude than the dipole and quadrupole correction energy discussed in section \ref{s:dipole}. Figure S10\textbf{a} uses the \textit{resampled} data set to show the dispersion energy difference $\Delta E_\mathrm{disp} = E_\mathrm{disp}^\mathrm{BG-adspecies} - E_\mathrm{disp}^\mathrm{BG-clean}$ between the BG systems with the adspecies *O, *OH, and *OOH and the clean system, all of which are in contact with water. This analysis therefore highlights the contribution of the dispersion energy to the adsorption energy for the solvated model systems. Figure 9\textbf{b} reproduces the \DDEsolv\ results as a function of the number of water molecules shown in figure \ref{f:ddesolv_resampled} but with the dispersion energy removed from the total energy.

This analysis shows that the dispersion contributions increase with the size of the solute. $\Delta E_\mathrm{disp}$ is close to zero for the *O adatom but \textit{ca.} -0.5 eV for *OOH in contact with 16 water molecules. The values for *OH and *OOH fluctuate significantly between subsequent data points, raising the question if the dispersion correction may be partially responsible for the erratic behavior of the \DDEsolv\ trends. However, analyzing the \DDEsolv\ trends in figure S10\textbf{b} shows that the results do not become more consistent when the dispersion energy contribution is removed. Hence, it can be concluded that any volatility of the dispersion correction results is also not the cause for but most likely the result of the erratic nature of the entire data set.

One caveat in this analysis and discussion, however, is that this \textit{a posteriori} removal of the final dispersion correction energy does not remove the entire influence of dispersion correction on the data set. Both the MD simulations and the local minimization of the structures in the \textit{resampled} data used dispersion correction throughout, hence the final structures (re-)analyzed here are generated on the RPBE-D3 potential surface. Despite this caveat, it is still unlikely that dispersion is the driving factor behind the erratic results since in particular the RPBE-D3 functional combination has been shown in the past to produce water structure that is in good agreement with experiments.\cite{tonigold2014}

\subsubsection{Influence of simulation cell size} \label{s:cellsize}

Simulation cells varied in size between simulations with different number of included water molecules. Because a PAW code was used and PAWs always fill the entire simulation cell, the $c$ cell parameter was minimized on a case-by-case basis to minimize the computational effort. Increasing or decreasing the box size also changes the total energy in a small way, hence it is important that all energy values used to calculate \DDEsolv\ in equation \eqref{eq:ddesolv_short} use the same cell dimensions. Consistency in this regard was ensured by generating the reference systems without solvent by removing water molecules from the original system; the reference systems are given alongside the solvated parent models in the data set available under DOI:10.5281/zenodo.7684918.

Furthermore, table S2 summarizes the total energy results for various reference systems without solvent from the MD data sets. The differences between system are, despite differences in the $c$ cell parameter, < 0.01 eV. Hence, the contribution from inconsistent cell dimensions, even if left untreated, are unlikely to distort results enough to account for the erratic results in this work.

\subsubsection{Influence of minimizing the reference systems}

This concern is related to the discussion about inconsistent cell size in section \ref{s:cellsize}. As pointed out there, the reference systems were obtained from the solvated parent systems by removal of the water molecules and subsequent energy-minimizaion of the resulting atomic configurations. This approach was chosen to account for the possibility that the most stable atomic arrangement of the BG-adspecies system may change once water molecules are removed. However, this approach creates a potential inconsistency: by optimizing the atomic configuration of the reference systems, the \DDEsolv\ values obtained from equation \eqref{eq:ddesolv_short} do not only contain the interaction of the BG-adspecies system with the water molecules but also the reorganization energy of the systems when going from a system in vacuum to a solvated system.

To investigate if energy minimization of the atomic configuration of the reference systems creates a bias, figure S11 compares \DDEsolv\ results from the \textit{one-shot optimization} data set where the reference systems without solvent were either minimized or where the reference energy contributions $E_{\mathrm{tot}}^{\mathrm{\,BG\ +\ adatom \ without\ solvent}}$ and $E_{\mathrm{tot}}^{\mathrm{\,BG\ without\ solvent}}$ were obtained from single-point total energy calculations. Results from this test show that the overall trends are identical. However, \DDEsolv\ for the adspecies in contact with 16, 24, and 32 water molecules are \textit{ca.}~0.2~eV more negative when obtained from single-point energy calculations based on the formerly-solvated atomic configurations. This result is unsurprising because the reference systems without water molecules can be assumed to be in a slightly unfavorable configuration when not allowed to relax under the new environmental conditions.

Overall, however, the differences appear to be systematic across the board and do not change the trends. Therefore, this factor is also not responsible for the erratic, non-converging behavior of \DDEsolv\ with increasing number of water molecules.

\subsubsection{Influence of using the lowest-energy structures to obtain the solvation stabilization energy}

\DDEsolv\ of the ORR intermediates on NG was calculated by Yu \textit{et al.} in 2011 by introducing 41 water molecules to a NG model, performing classical dynamics simulations with DFT fores, and finally minimizing the lowest-energy solvated structures obtained from the MD simulation with respect to the atomic coordinates.\cite{Yu2011} The group obtained \DDEsolv\ values of -0.53, -0.38, and -0.49 eV for the *O, *OH, and *OOH intermediates, respectively. While this approach fails to capture the vast structural diversity accessible to the system and is therefore less representative of the system under experimental conditions, it has value from a computational perspective because \DDEsolv\ according to equation \ref{eq:ddesolv_short} is calculated exclusively from 4 values total, all of which represent the best possible guess for the global minimum energy configuration of each system.

Hence, this approach is applied to the present data set. The \textit{flexible MD} data set was re-analyzed to find the structure with the lowest total energy for each combination of adspecies and number of water molecules. The obtained images were then energy-minimized using the tight accuracy settings outlined in section \ref{s:computational}. Figure S12 shows the results of this approach.

Figure S12 shows that the \DDEsolv\ results for *OH and *OOH are somewhat comparable to the \textit{resampled} data set, which is most closely related to this test, in terms of relative trends but less so in terms of absolute values. However, the *O intermediate shows significantly more negative \DDEsolv\ results.

It can be concluded that this approach not only did not resolve the erratic results but can further distort the results because the close-to-ideal local configurations optimized in this case likely do not represent the average configurations of water molecules around the adspecies in real, finite-temperature systems.

\subsubsection{Influence of constraining the geometry of the BG sheet}

100 ps of classical dynamics without bond length constraints on the water molecules and \textit{no} geometry constraint on the BG sheet and adspecies were accidentally performed for the BG-OOH system in contact with 1 layer of water. This mistake, however, can be used to probe the influence of the geometry constraint on the BG-OOH system.

Figure S13 compares the total energy and temperature trends over the course of the simulation time for the simulations with and without geometry constraint on the BG-OOH system. Most notably, the total energy fluctuations are significantly increased in the case of the model without constraint. The increased amplitude of fluctuations translate to a larger error bar. Hence, without the geometry constraint on the BG-OOH backbone, more sampling statistics is required to reduce the uncertainty to an appropriate level. In the interest of computational feasibility, the geometry constraint therefore turns out to be an almost necessary prerequisite.

Finally, figure S14 compares the $g(z)$ of the systems where the BG sheet was constrained against that of the non-constrained system. No significant differences were observed. This result indicates that constraining the BG sheet does not significantly affect the interactions between the surface and the first water layer from a structural point of view.

\subsubsection{Embedded solvation approach}

The embedded solvation approach, where a small cluster of explicit solvent molecules is used in combination with an implicit continuum description of the solvent bulk, has recently been employed to good effect.\cite{GarciaSOLV,vandenbossche2019} In the beginning of this study, the \textit{one-shot minimization} data set was, in fact, computed using the embedded approach and similarly erratic results were obtained. The implicit solvent model was then discarded for the remainder of this study to reduce the number of potential error sources.

\section{Conclusion}\label{s:conclusion}

Density functional theory-driven minimization calculations and classical molecular dynamics simulations were used to obtain the solvation stabilization energy, \DDEsolv, for the oxygen reduction reaction intermediates *O, *OH, and *OOH adsorbed on a Boron-doped graphene sheet in contact with 8, 16, 24, and 32 molecules of water. The goal of this study was to apply the obtained \DDEsolv\ values to accurate hybrid DFT adsorption energy results for the ORR intermediates to refine potential-dependent free-energy predictions by including the influence of solvation. Although 4 different data set were obtained that sampled \DDEsolv\ from the model systems in different ways using static and dynamic calculations, no converged \DDEsolv\ result were obtained.

A detailed discussion of the simulation parameters and potential error sources is provided to rule out that technical errors lead to these erratic results. We conclude that 32 water molecules, which is the equivalent of 4 layers of water in this model system, are not sufficient to describe solvation of the adspecies within chemical accuracy. Chemical accuracy, \textit{i.e.} convergence of \DDEsolv\ to changes of < 0.05~eV when adding more and more water molecules, is essential since any reduction of the free energy of the potential-determining intermediate will lead to a proportional reduction of the thermochemical overpotential as well.

These results emphasize that new simulation methods are required to be able to calculate large enough systems to obtain converged \DDEsolv\ results since molecular dynamics simulations with DFT forces quickly become computationally unfeasible when adding more and more water molecules. Our groups are therefore focused on implementing a 2D periodic hybrid method (often referred to as QM/MM) for the open-source ASE and GPAW software packages which will enable calculations with thousands of water molecules.

Another promising approach to tackle this problem is the recently developed on-the-fly machine learning force field training method.\cite{Jinnouchi2020} This approach could be used to train a machine learning force field on a small system and then upscale the system to contain many water molecules while retaining close-to-DFT accuracy.

Finally, we believe in the importance of presenting these negative results to the catalysis community as a word of caution. It is easy to underestimate the number of explicit water molecules required to obtain sufficiently accurate solvation energy results.

\backmatter

\bmhead{Supplementary information}

A Supplementary Information document is available online. The simulation results and input files are available via DOI:10.5281/zenodo.7684918. The data analysis procedure for all tables and figures in the main manuscript and supplementary information file is available via \href{https://bjk24.gitlab.io/bg-solvation/}{https://bjk24.gitlab.io/bg-solvation/}.

\bmhead{Acknowledgments}

This work was supported in part by the Icelandic Research Fund (grant no. 174582-053).

\bmhead{Author Contributions}

All authors contributed to the study conception and design. Simulations, data collection and analysis were performed by BK. The first draft of the manuscript was written by BK and all authors commented on previous versions of the manuscript. All authors read and approved the final manuscript.

\bmhead{Compliance with Ethical Standards}

The authors declare no competing interests.


\bibliography{sn-article.bib}


\setcounter{figure}{0}
\setcounter{table}{0}
\renewcommand{\thefigure}{S\arabic{figure}}
\renewcommand{\thetable}{S\arabic{table}}

\section{Supplementary Information}

\subsection{The model system}

\begin{figure}[htbp]
    \centering
    \begin{minipage}[t]{0.05\linewidth}
        \vspace{0pt}
        \textbf{a}
    \end{minipage}
    \begin{minipage}[t]{0.4\linewidth}
        \vspace{0pt}
        \includegraphics[width=\linewidth]{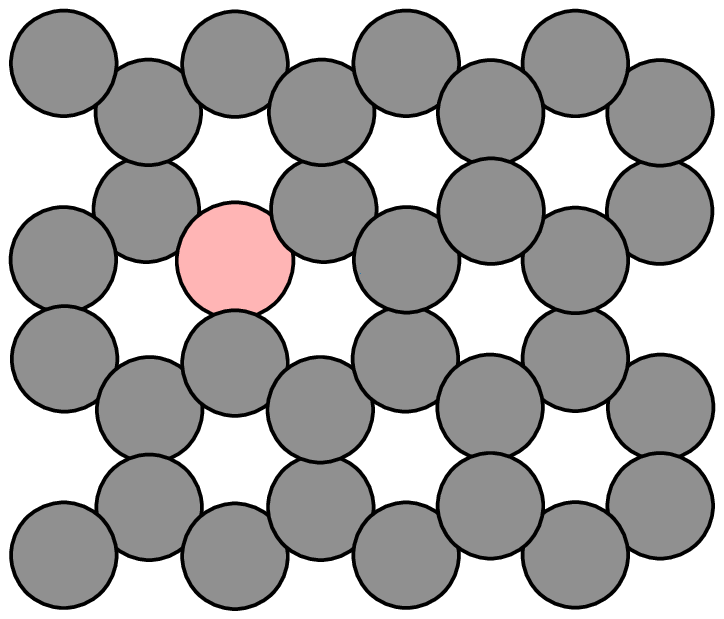}
    \end{minipage}
    \begin{minipage}[t]{0.05\linewidth}
        \vspace{0pt}
        \textbf{b}
    \end{minipage}
    \begin{minipage}[t]{0.4\linewidth}
        \vspace{0pt}
        \includegraphics[width=\linewidth]{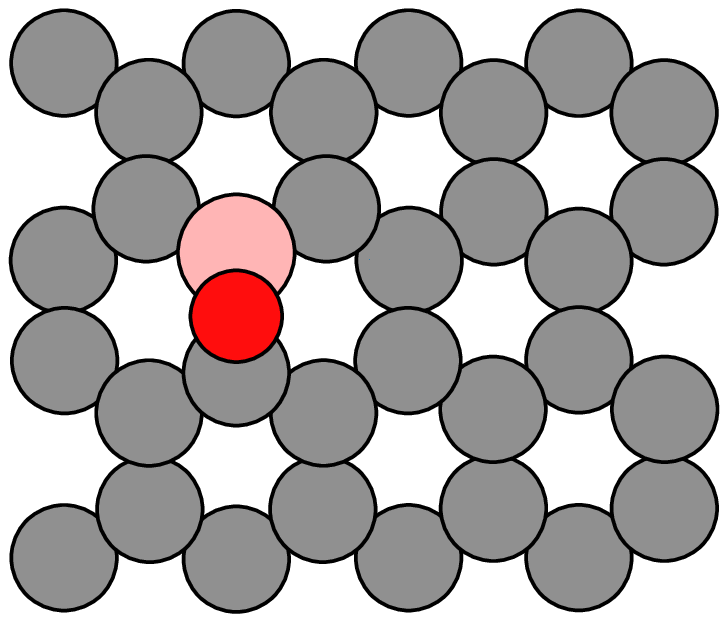}
    \end{minipage}\\
    \begin{minipage}[t]{0.05\linewidth}
        \vspace{0pt}
        \textbf{c}
    \end{minipage}
    \begin{minipage}[t]{0.4\linewidth}
        \vspace{0pt}
        \includegraphics[width=\linewidth]{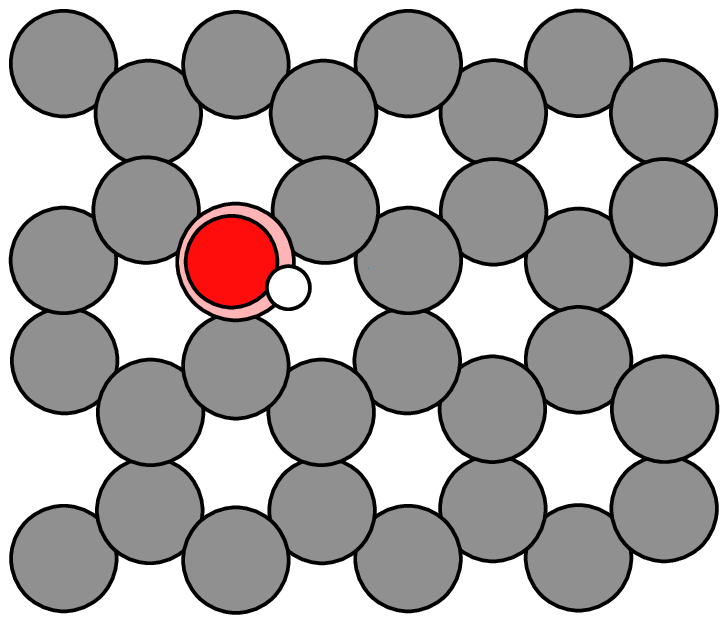}
    \end{minipage}
    \begin{minipage}[t]{0.05\linewidth}
        \vspace{0pt}
        \textbf{d}
    \end{minipage}
    \begin{minipage}[t]{0.4\linewidth}
        \vspace{0pt}
        \includegraphics[width=\linewidth]{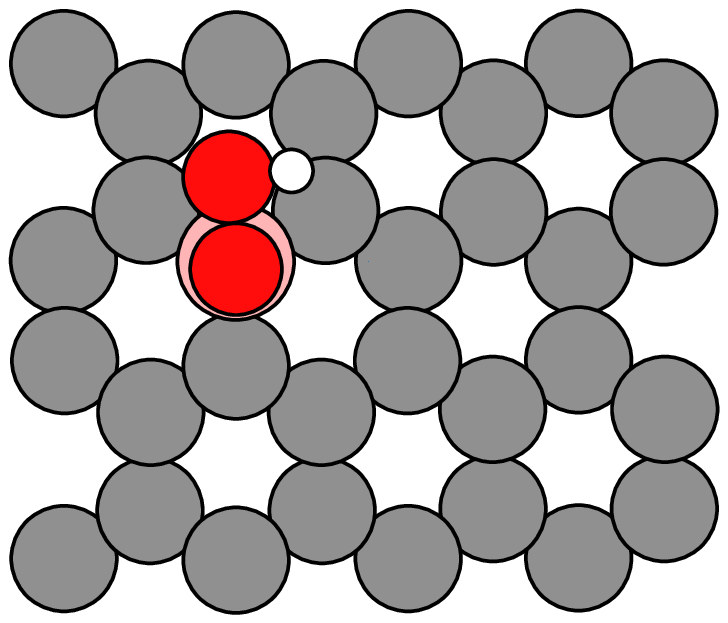}
    \end{minipage}\\

    \caption{Illustrations of the BG sheet model (\textbf{a}) in contact with *O (\textbf{b}), *OH (\textbf{c}), and *OOH (\textbf{d}) adspecies.}
    \label{f:models_vac}
\end{figure}

\begin{figure}[htbp]
    \centering
    \begin{minipage}[t]{0.05\linewidth}
        \vspace{0pt}
        \textbf{a}
    \end{minipage}
    \begin{minipage}[t]{0.4\linewidth}
        \vspace{0pt}
        \includegraphics[width=\linewidth]{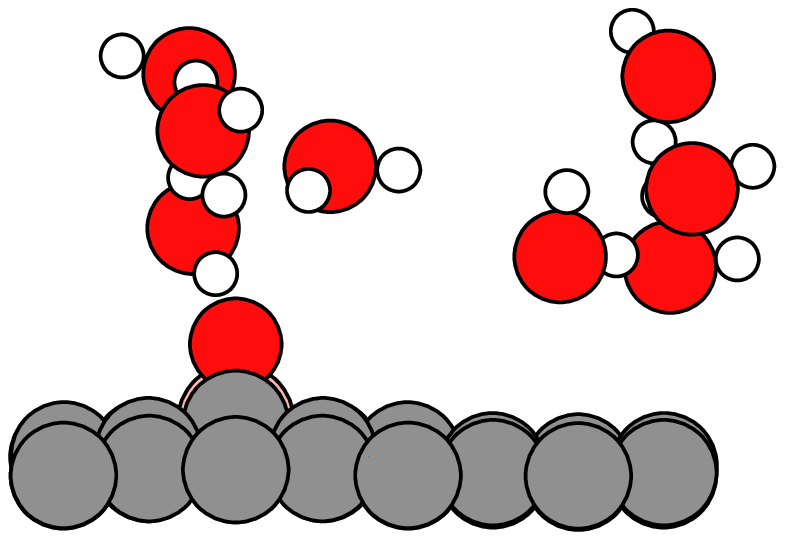}
    \end{minipage}
    \begin{minipage}[t]{0.05\linewidth}
        \vspace{0pt}
        \textbf{b}
    \end{minipage}
    \begin{minipage}[t]{0.4\linewidth}
        \vspace{0pt}
        \includegraphics[width=\linewidth]{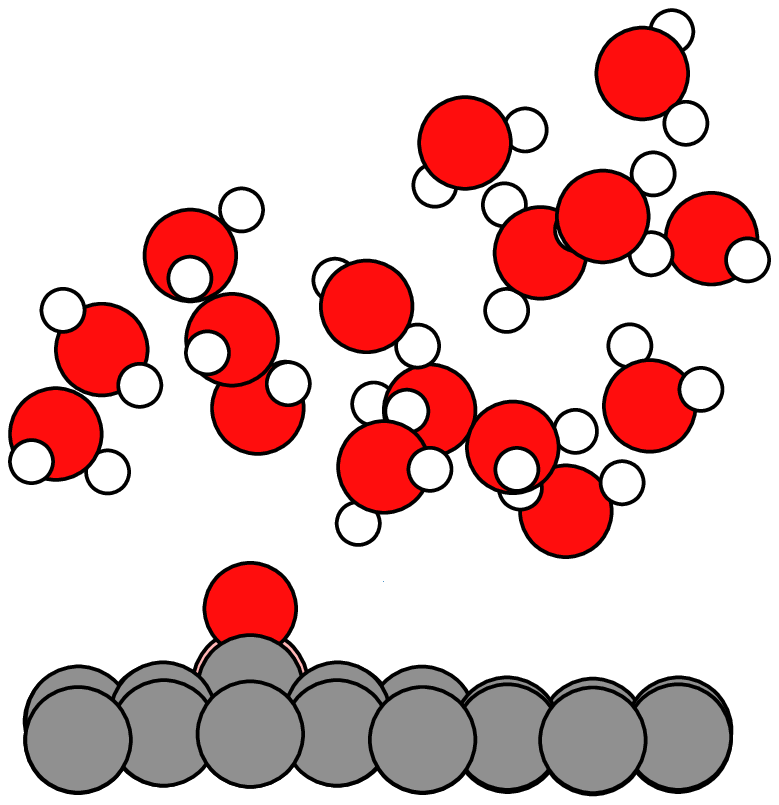}
    \end{minipage}\\
    \begin{minipage}[t]{0.05\linewidth}
        \vspace{0pt}
        \textbf{c}
    \end{minipage}
    \begin{minipage}[t]{0.4\linewidth}
        \vspace{0pt}
        \includegraphics[width=\linewidth]{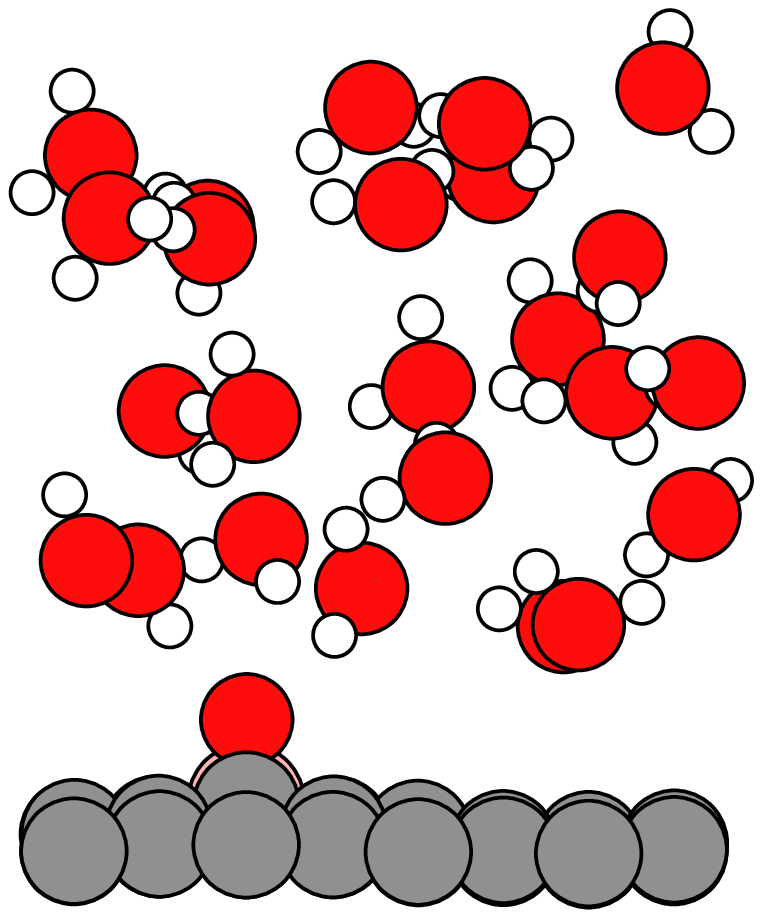}
    \end{minipage}
    \begin{minipage}[t]{0.05\linewidth}
        \vspace{0pt}
        \textbf{d}
    \end{minipage}
    \begin{minipage}[t]{0.4\linewidth}
        \vspace{0pt}
        \includegraphics[width=\linewidth]{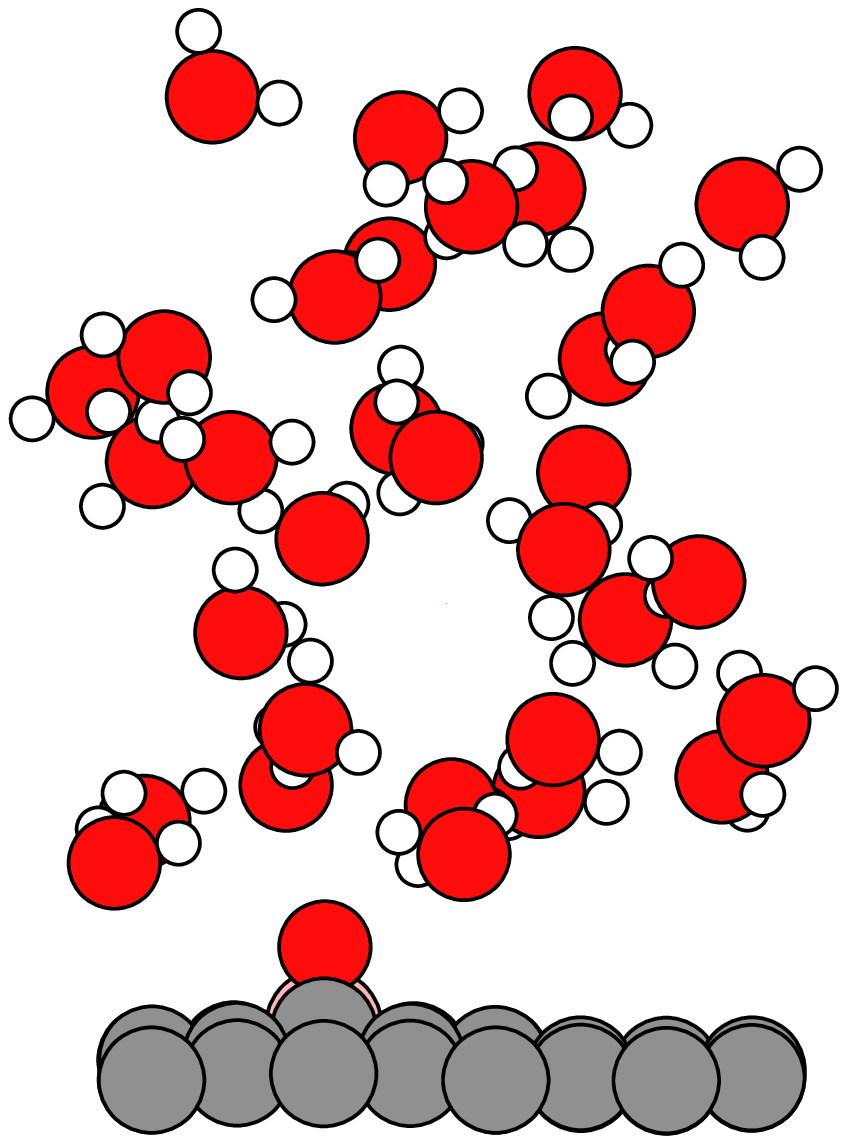}
    \end{minipage}\\

    \caption{Illustrations of the BG sheet model with an *O adatom in contact with 1 (\textbf{a}), 2 (\textbf{b}), 3 (\textbf{c}), and 4 layers (\textbf{d}) of water molecules.}
    \label{f:models_solv}
\end{figure}

Figure \ref{f:models_vac} illustrates the 32-atomic BG sheet model with the *O, *OH, and *OOH adspecies used throughout this study. Figure \ref{f:models_solv} illustrates the BG sheet model with an *O adatom in contact with 1-4 layers of water molecules. Interactive visualizations of the model systems can be found online at \url{https://bjk24.gitlab.io/bg-solvation/docs/visualization.html}.

\subsection{Density functional benchmark and ORR free energy trends} \label{s:benchmark}

\begin{figure}[htbp]
	\centering
    \includegraphics[width=\linewidth]{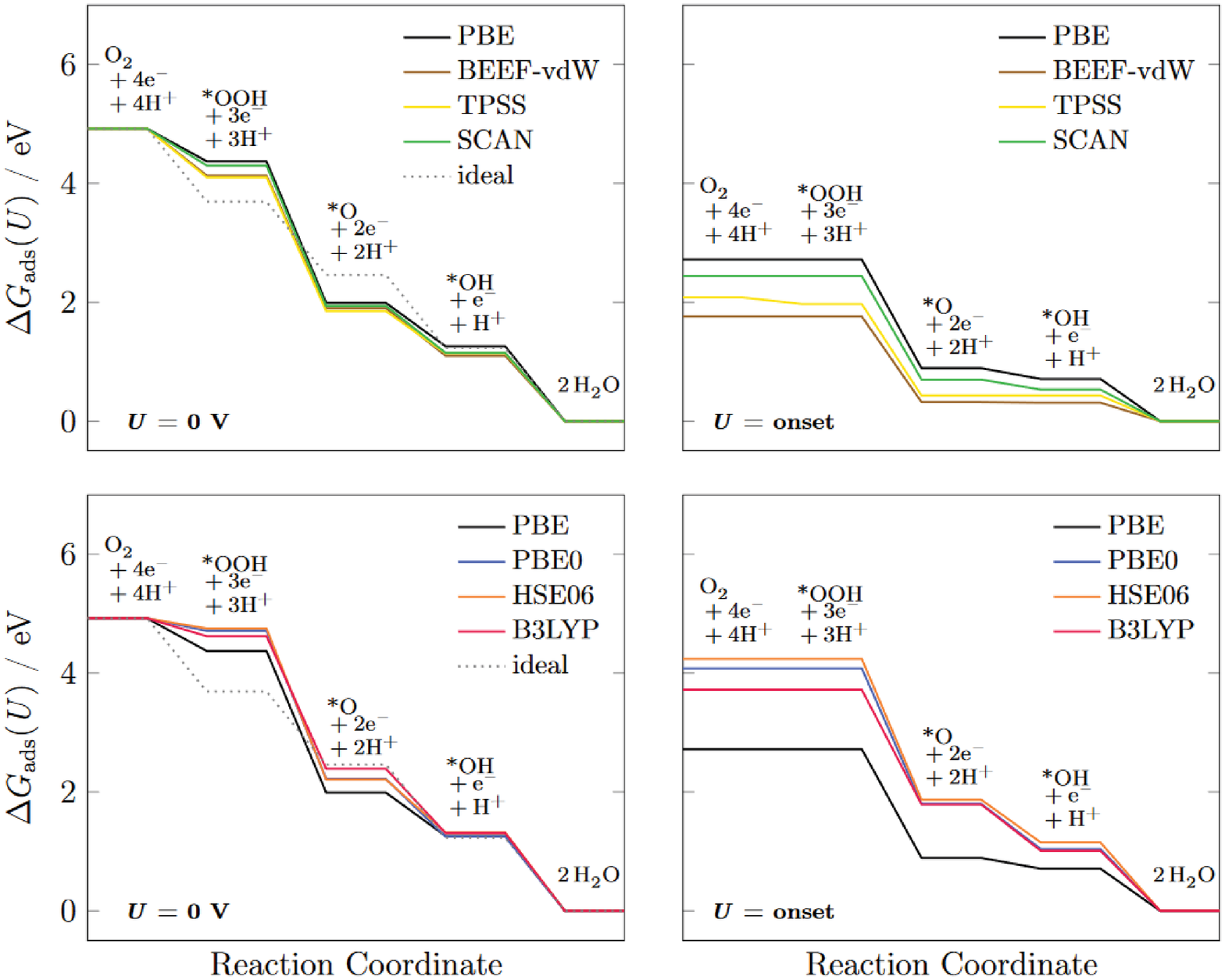}
    \caption{Free energy diagrams for the 32-atomic BG model at 0~V obtained using GGA and meta-GGA functionals (top) as well as PBE and hybrid functionals (bottom). A hypothetical ideal ORR catalyst with a free energy change of 1.23~V at each reaction step is shown as a dotted line.}
	\label{fgr:BG_gibbs}
\end{figure}

Adsorption free energy values for the ORR intermediates *O, *OH, and *OOH and, from these, thermochemical overpotentials \tcm\ are calculated for the 32-atomic BG model using various different density functionals using the computational hydrogen electrode free energy method.\cite{norskov2004} Potential-dependent free energy values are calculated the same way as in an earlier work on NG.\cite{Kirchhoff2021a} Using \tcm\ as a descriptor for ORR activity, this test is performed to investigate how strongly the thermochemical results depend on the chosen density functional. In a previous study on NG, the HSE06 functional was found to reproduce a Diffusion Monte Carlo benchmark value most accurately out of all tested functionals.\cite{Kirchhoff2021a} Thus, the HSE06 result is used as a reference value in the following. Figure \ref{fgr:BG_gibbs} shows free energy changes of the ORR intermediates on the BG model at 0~V~\textit{vs.}~SHE and at the extrapolated onset potential for each functional.

The potential determining step (PDS) in all cases is the formation of the *OOH intermediate. Compared to a hypothetical ideal ORR catalyst with a free energy change of 1.23~V at each reaction step, the *OOH intermediate is underbound in the case of all functionals, with hybrid functionals underbinding *OOH more strongly than meta-GGA and GGA functionals. The *O intermediate is significantly overbound in the case of GGA and meta-GGA functionals which is in agreement with benchmark calculations on undoped graphene and free energy calculations on NG.\cite{hsing2012,janesko2013,Kirchhoff2021a} The energetic description of the *OH intermediate is similar for all tested functionals and aligns well with the hypothetical ideal catalyst. The free energy results highlight a significant issue: all functionals are in good agreement for the *OH intermediate but differ in their results for the *OOH and *O adspecies. Therefore, different functionals are bound to produce different overall trends.

To further illustrate this conclusion, thermochemical overpotentials \tcm\ are calculated from the extrapolated onset potentials $U_\text{onset}$ as
\begin{equation}
    \text{\tcm} = 1.23~\text{V} - U_\text{onset}.
\end{equation}
\tcm\ results are summarized in Table \ref{tab:overpotentials}.
\begin{table}[htbp]
	\caption{Thermochemical overpotentials \tcm\ obtained for the 32-atomic BG model with various density functionals. $^*$ Largest and smallest possible \tcm\ obtained using standard deviations of the adsorption free energy values of the ORR intermediates, based on Bayesian error estimation using an ensemble of 2000 functionals.}
	\label{tab:overpotentials}
    \centering
    \renewcommand{\arraystretch}{1.5}
    \begin{tabular}{c|c}
         Density Functional & \tcm\ / V \\ \hline
         HSE06 & 1.06 \\
         PBE0 & 1.02 \\
         B3LYP & 0.93 \\
         PBE & 0.68 \\
         SCAN &  0.61 \\
         TPSS & 0.52 \\
         BEEF-vdW & 0.44 \\
         & (0.26--0.80)$^*$
    \end{tabular}
\end{table}
The hybrid functionals HSE06 and PBE0, which constitute the most reliable result according to benchmarking,\cite{Kirchhoff2021a} perform similarly and give the highest \tcm\ out of all tested functionals with 1.06 and 1.02~V, respectively. The B3LYP functional produces a slightly lower \tcm\ of 0.93~V. The tested GGA and meta-GGA functionals give significantly lower \tcm\ values of 0.68~V (PBE), 0.61~V (SCAN), 0.52~V (TPSS), and 0.44~V (BEEF-vdW). These trends are analogous to previous computational results for NG.\cite{Kirchhoff2021a} Notably, choosing a meta-GGA functional will not provide significant improvements over GGAs.

Bayesian error estimation is performed based on an ensemble of 2000 functionals generated by BEEF-vdW to obtain standard deviations for the adsorption free energy values of the ORR intermediates. Based on the error estimation, the largest- and smallest-possible \tcm\ value can be calculated which should be indicative of the overall uncertainty of GGA functionals for this application. The \tcm\ range obtained this way for BEEF-vdW is 0.26--0.80~V. Notably, this range does not include values obtained by hybrid functionals. Since the HSE06 and PBE0 hybrid functionals are able to reproduce DMC benchmark values for graphene-based materials,\cite{Kirchhoff2021a, hsing2012, janesko2013} this result indicates that GGA functionals lack some fundamental contribution - likely exact exchange - that is necessary to accurately describe the electronic structure of this material class.

\subsection{Convergence studies}

\subsubsection{Supercell size convergence}

A supercell size convergence study is performed with the generalized gradient approximation (GGA) functional by Perdew, Burke, and Ernzerhof (PBE) as well as with the hybrid functional by Heyd, Scuseria, and Ernzerhof (HSE06) where the calculation of exact exchange was downsampled:
\begin{enumerate}
    \item PBE-based optimization (labeled "PBE")
    \item HSE06-based optimization where the \textit{k} grid was reduced to the $\Gamma$ point for the Hartree-Fock portion of the calculation (labeled "HSE06-fast")
\end{enumerate}
The downsampling of the HSE06 functional reduces the \textit{k} grid to the $\Gamma$ point for the calculation of the Hartree-Fock exchange energy, thereby making minimization of the atomic coordinates computationally feasible at the hybrid DFT level.

\begin{figure} [htbp]
    \centering
    \includegraphics[width=0.8\linewidth]{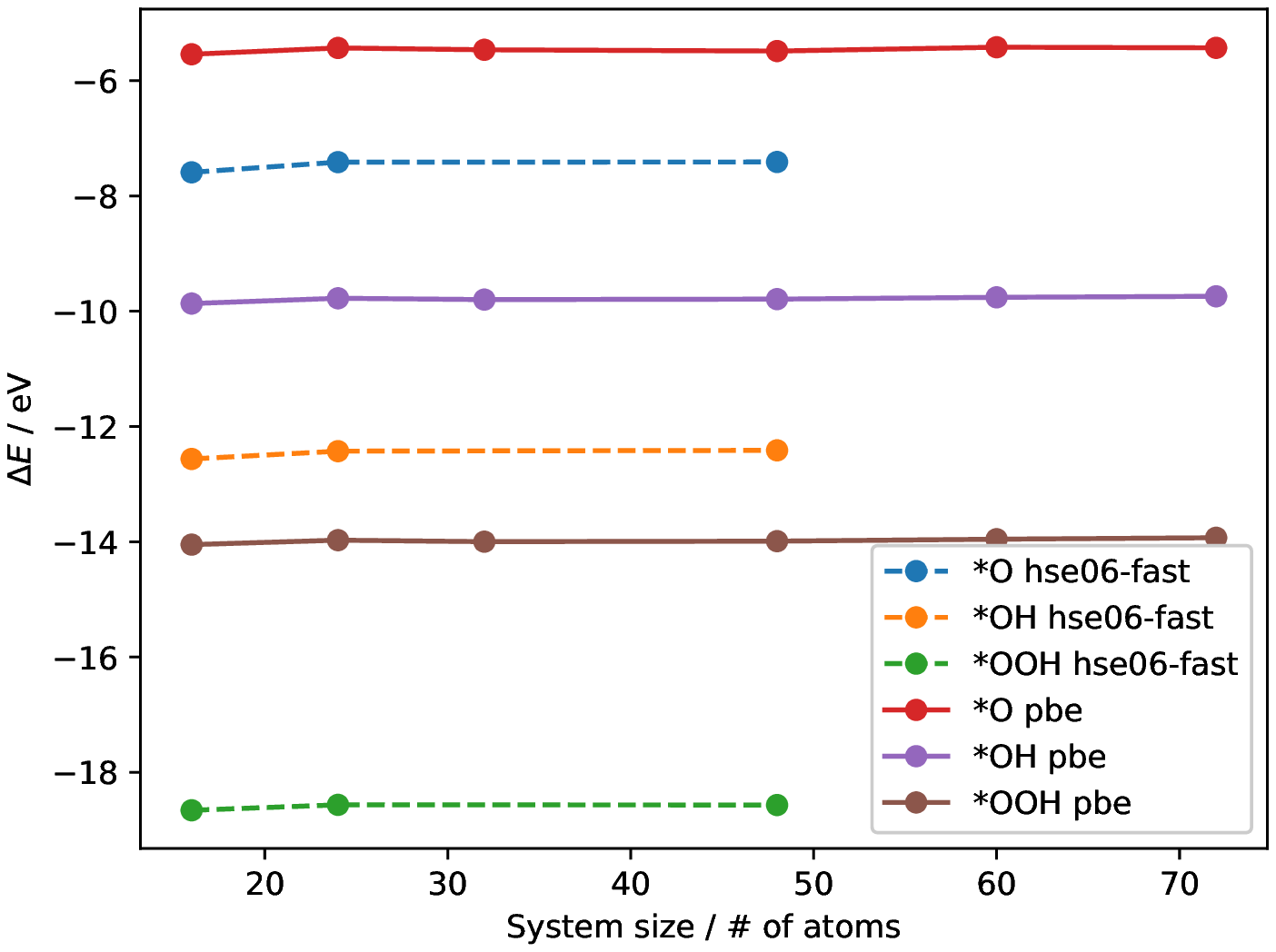}
    \caption{Supercell size convergence study where $\Delta E$ is calculated according to equation \eqref{eq:ediff} for the BG sheet in contact with O, OH, and OOH adspecies using the PBE and a downsampled HSE06 functional (see text for details).}
    \label{f:supercell}
\end{figure}

The test below calculates energy differences according to
\begin{equation}
    \Delta E = E^\mathrm{ads}_\mathrm{tot} - E^\mathrm{clean}_\mathrm{tot}, \label{eq:ediff}
\end{equation}
where $E^\mathrm{ads}_\mathrm{tot}$ is the total energy of a system with an adspecies (*O, *OH, or *OOH) and $E^\mathrm{clean}_\mathrm{tot}$ is the total energy of the BG sheet without any adspecies. To test for supercell size convergence, it is not necessary to calculate an actual adsorption energy by taking into account the total energy of the adspecies calculated from molecules such as O$_2$, H$_2$, and H$_2$O because their total energy will change only in small ways as the size of the supercell increases. There is always a slight change because plane waves always fill the entire box, also for molecules, which makes the total energy dependent on the box size. However, this effect is insignificant compared to the influence of decreasing dopant and adatom concentration as a function of the supercell size. Figure \ref{f:supercell} shows the results of this test.

Since these convergence tests were performed at a much earlier date than the production calculations shown in the Results section of the main manuscript, slightly different settings were used in this test compared to the settings summarized in the Computational Details section of the main manuscript. The following list addresses the differences and their potential impact on the results:

\begin{enumerate}
    \item This data set uses a PAW energy cutoff of ENCUT = 500. As can be seen further below in section \ref{s:cutoff}, the choice of ENCUT = 600, which was used in the main article for geometry optimization calculations, is on the paranoid end of safe and there is no reason to assume that using a cutoff of 500 eV in this instance had an adverse effect on the results.

    \item Note that the \textit{k} grid was individually optimized for each system. The \textit{k} grid optimization runs are not shown for the sake of brevity. The entire calculated data set is provided in a archive found under DOI:10.5281/zenodo.7684918; consult the KPOINTS files in the respective subfolders for the converged \textit{k} grid settings.

    \item We expect that the trends obtained from these functionals (PBE and HSE06, the latter of which constitutes a PBE functional with 25\% exact exchange and a screening parameter) are fully transferable to the RPBE functional with DFT-D3 dispersion correction that was used for production calculations later on. RPBE and PBE belong to the same family of GGA-rung functionals, more specifically RPBE is a re-parameterized PBE functional optimized towards surface (adsorption) calculations. The differences between PBE and RPBE are minor, unlike for example the differences between PBE and functionals like BEEF or SCAN which use fundamentally different potential terms and would therefore require careful re-investigation.

    \item  The influence of DFT-D3 on the $\Delta E$ results is negligible because the adspecies are covalently bound. DFT-D3 is therefore unlikely to affect the results of this kind of convergence test but will become more important as layers of non-covalently bound water molecules are added to the model systems. The influence of DFT-D3 on the results is investigated in more detail in the Discussion section.
\end{enumerate}

This convergence test shows that the energy differences are well converged from the beginning. This result is in contrast to what was observed for NG where the the size 16 and 24 data points did not show converged results yet.\cite{Kirchhoff2021a} However, to stay consistent with the previous work on NG, we chose to use the 32 atomic model system going forward.

There is also another reason for using the slightly larger system: there is the possibility that if the system size is chosen too small, the water atoms in the individual layers become too crowded and do not have the necessary space to relax and accommodate the surface and adspecies properly. To investigate crowding effects in the lateral directions properly, the MD simulations presented later on in section Results should be repeated for a set of surface models with increasing size in $x$ and $y$ direction; however, such a test was not computationally feasible at the time of this study.

\subsubsection{\textit{k} grid convergence of the solvation stabilization energy}

Convergence of $\Delta \Delta E_\mathrm{solv}$ with respect to the \textit{k} grid is tested. Our hypothesis was that the \textit{k} grid density of $\Delta E_\mathrm{ads}$ and $\Delta \Delta E_\mathrm{solv}$ should be different because the interaction is fundamentally different (covalent interaction of adatom with periodic surface \textit{vs.} non-convalent long-range interaction of solvent molecules with - mostly - the adspecies and only very lightly with the hydrophobic graphene surface). This hypothesis was strengthened by the observation that  $\Delta E_\mathrm{ads}$ and $\Delta \Delta E_\mathrm{solv}$ do not share the same dependence on the density functional.\cite{Kirchhoff2021a}

\begin{figure}[htbp]
    \centering
    \includegraphics[width=0.8\linewidth]{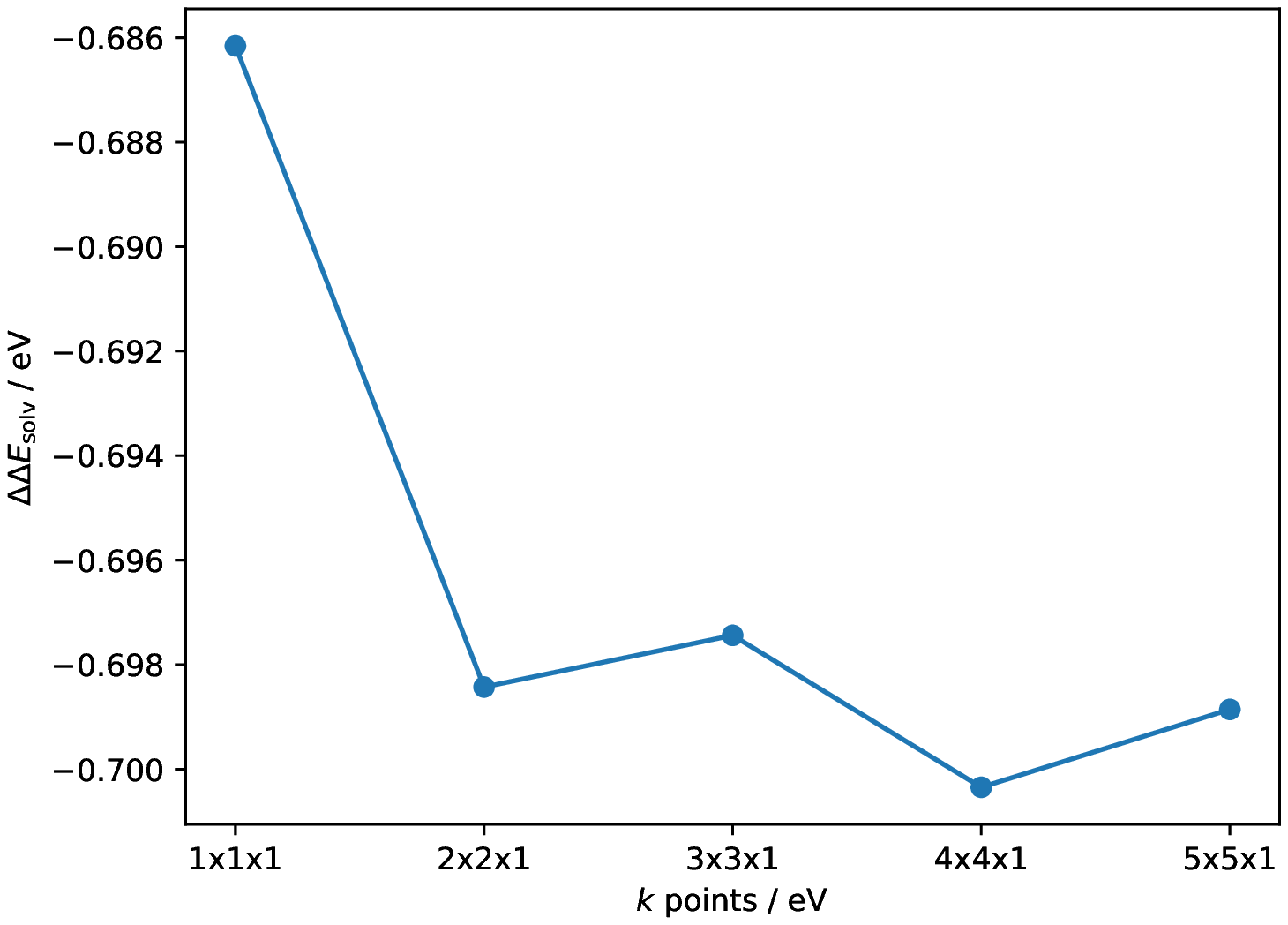}
    \caption{Convergence study of $\Delta \Delta E_\mathrm{solv}$ with respect to the \textit{k} grid, where $\Delta \Delta E_\mathrm{solv}$ is calculated according to equation \eqref{eq:ddesolv_conv} for the BG sheet in contact with an O adatom using the PBE functional.}
    \label{f:kgrid}
\end{figure}

In this case $\Delta \Delta E_\mathrm{solv}^\mathrm{O*}$ for the BG system with an *O adatom is calculated as
\begin{equation}
    \Delta \Delta E_\mathrm{solv}^\mathrm{\,*O} = E_\mathrm{tot}^\mathrm{\,BG+O+solv} - E_\mathrm{tot}^\mathrm{\,BG+O} - E_\mathrm{tot}^\mathrm{\,solv}, \label{eq:ddesolv_conv}
\end{equation}
where $E_\mathrm{tot}^\mathrm{\,BG+O+solv}$ is the total energy of the BG sheet model with the adatom and an overlayer of 8 H$_2$O molecules, $E_\mathrm{tot}^\mathrm{\,BG+O}$ is the total energy of the BG sheet model with the adatom without any solvent molecules, and $E_\mathrm{tot}^\mathrm{\,solv}$ is the total energy of only the 8 H$_2$O molecules. All total energy values are obtained as single-point results from the same starting geometry; the atomic positions are not optimized for the different subsystems. Figure \ref{f:kgrid} shows the results of this test.

From this test, $\Delta \Delta E_\mathrm{solv}$ results are converged using a 2x2x1 \textit{k} grid. Arguably, the results can be regarded converged already at a 1x1x1 grid since the energy difference between the smallest and the next \textit{k} grid is only \textit{ca.} 0.012 eV. For the MD simulations, we erred on the side of caution and used a 3x3x1 \textit{k} grid. Static calculations used various \textit{k} grids depending on the exact systems; consult the data set under DOI:10.5281/zenodo.7684918 for exact \textit{k} grid settings for each  subset of simulations.

\subsubsection{Convergence of the plane-augmented wave energy cutoff} \label{s:cutoff}

\begin{figure}[htbp]
    \centering
    \includegraphics[width=0.8\linewidth]{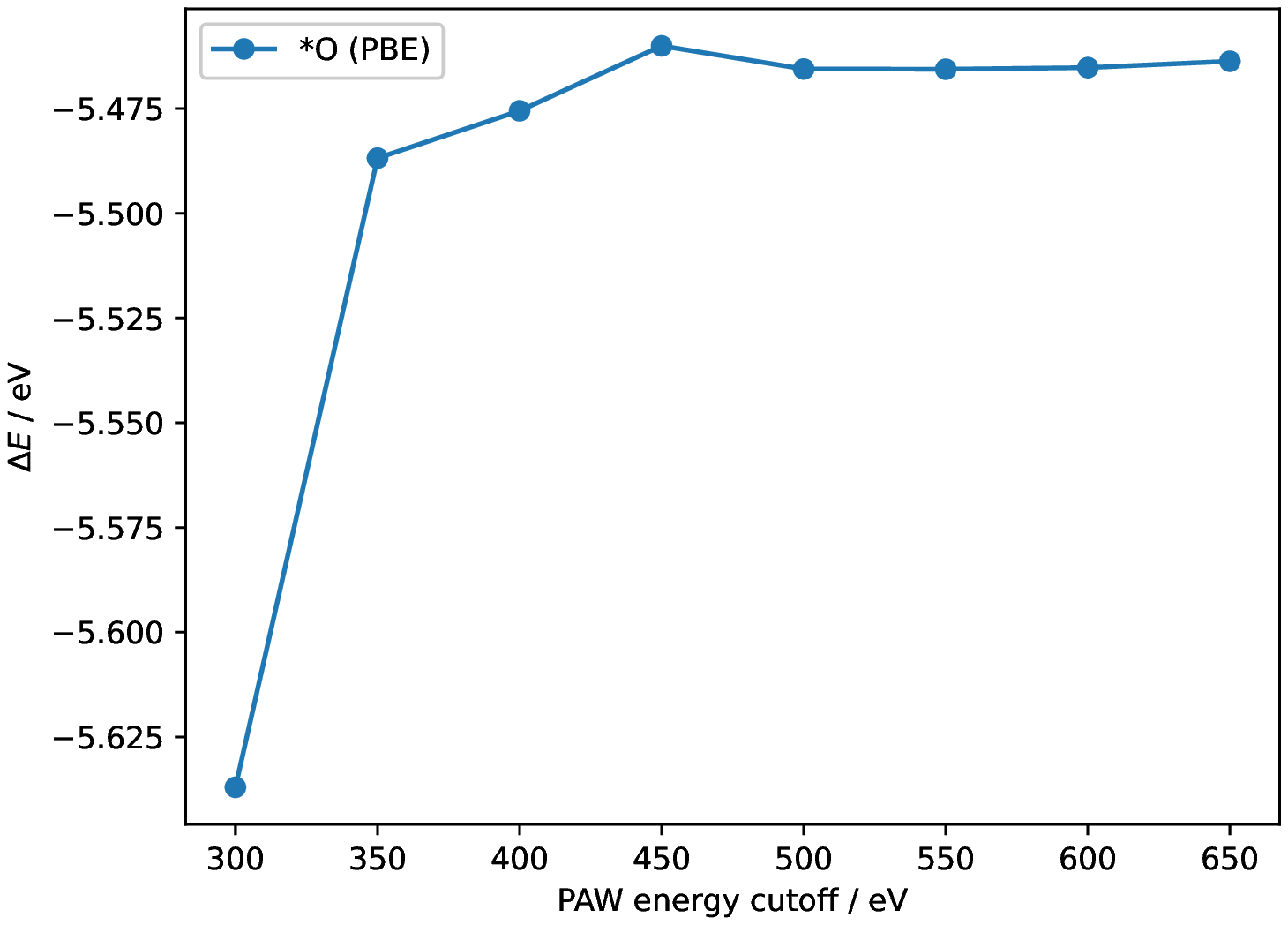}
    \caption{Plane-augmented wave energy cutoff convergence study where $\Delta E$ is calculated according to equation \eqref{eq:ediff} for the BG sheet in contact with an O adatom using the PBE functional.}
    \label{f:encut}
\end{figure}

The same approach as above for the supercell size was used to establish  the relationship between $\Delta E$ and the plane-augmented wave energy cutoff. Only the *O adatom is tested in this case since behavior appears to be similar for all intermediates (see for example figure \ref{f:supercell}) and because the *O intermediate in particular was found to be the most notorious in benchmark calculations with NG.\cite{Kirchhoff2021a} The 32-atomic BG sheet model was used. Aside from using the PBE functional and a changing ENCUT parameter, the other simulation parameters were consistent with those summarized in the Computational Details section in the main article. Figure \ref{f:encut} shows the results of this test.

Results show that a PAW cutoff energy of 500 eV is sufficient to obtain converged relative energy results. For geometry optimization calculations, a cutoff of 600 eV was chosen in an attempt to err on the side of caution. However, MD simulations were performed with ENCUT = 400 due to the excessive computational cost of higher energy cutoff values. See section Discussion in the main article for an analysis of this disparity.

As above for the supercell size convergence, there is no reason to assume that this convergence test with the PBE functional would not translate to the RPBE + DFT-D3 functional combination later on as they are closely related functionals from the GGA-DFT rung.

\subsection{Oxygen \textit{z} distribution ($g(z)$) results}

\begin{figure}[htbp]
    \centering
    \includegraphics[width=0.49\linewidth]{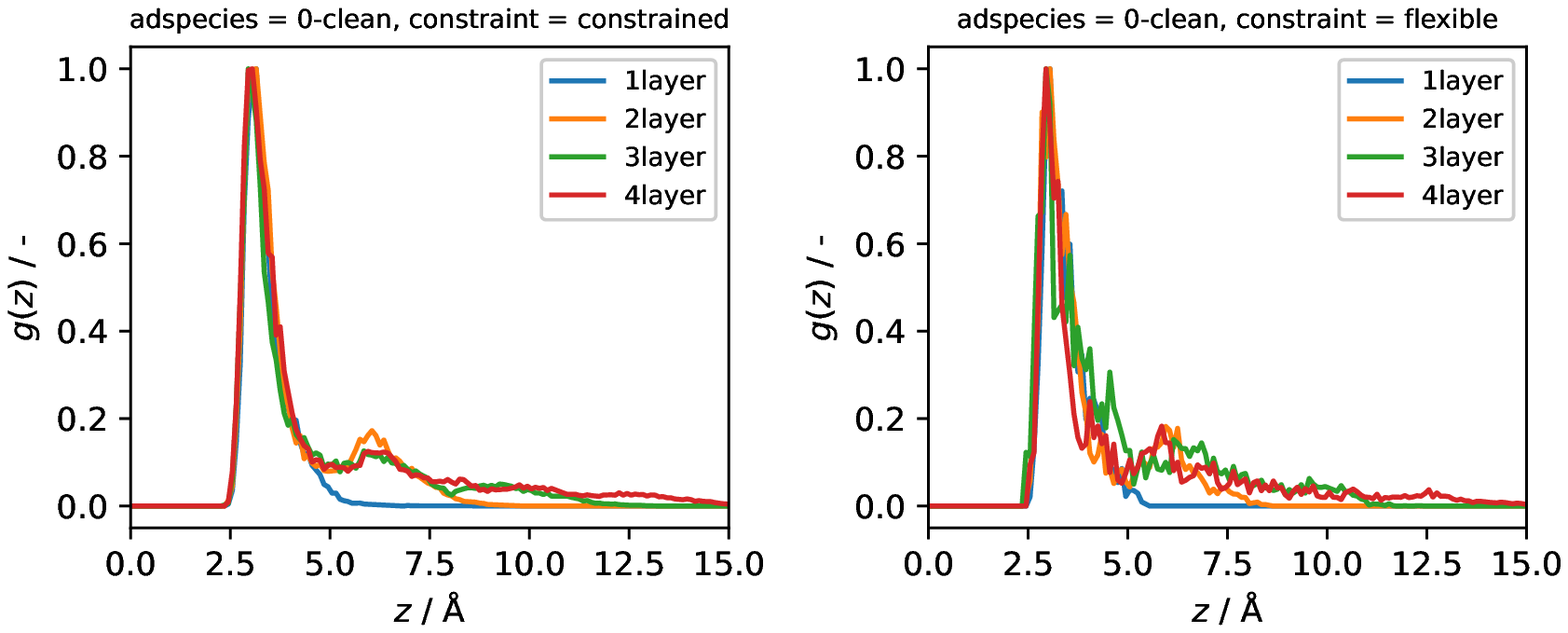}
    \includegraphics[width=0.49\linewidth]{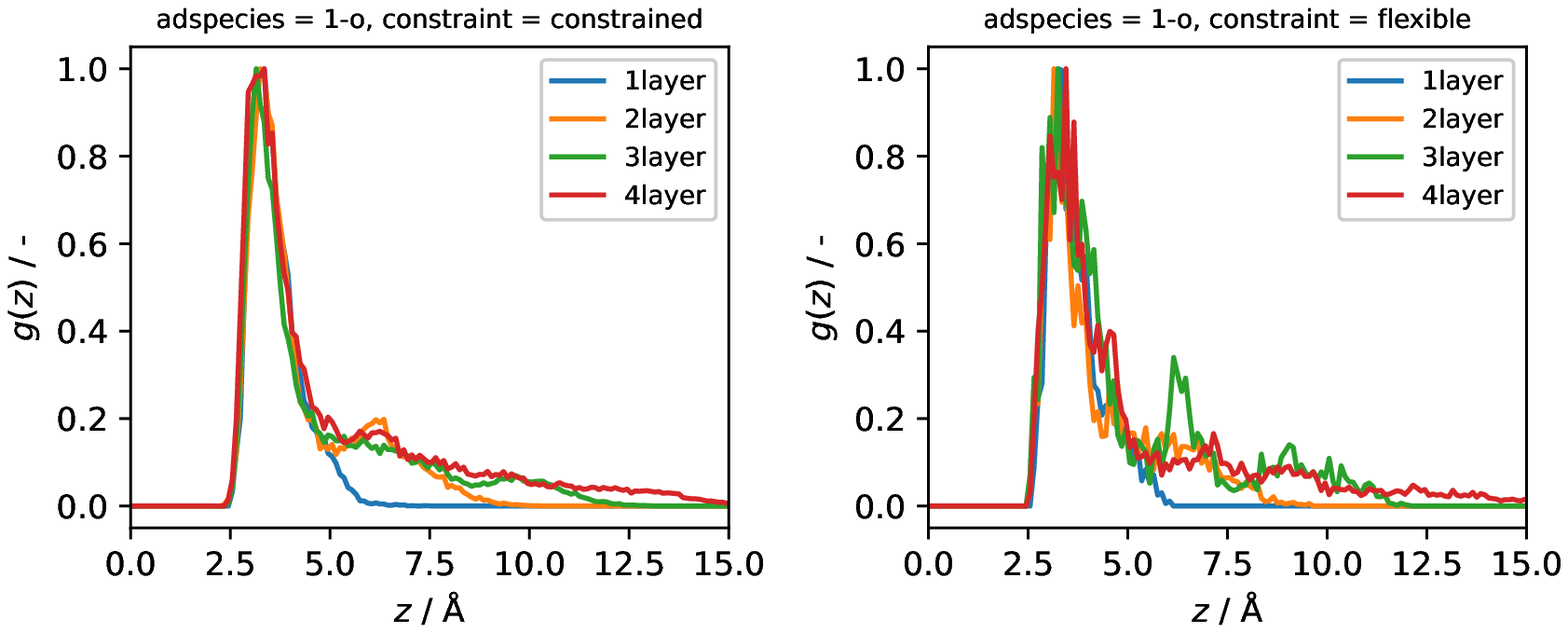}\\
    \includegraphics[width=0.49\linewidth]{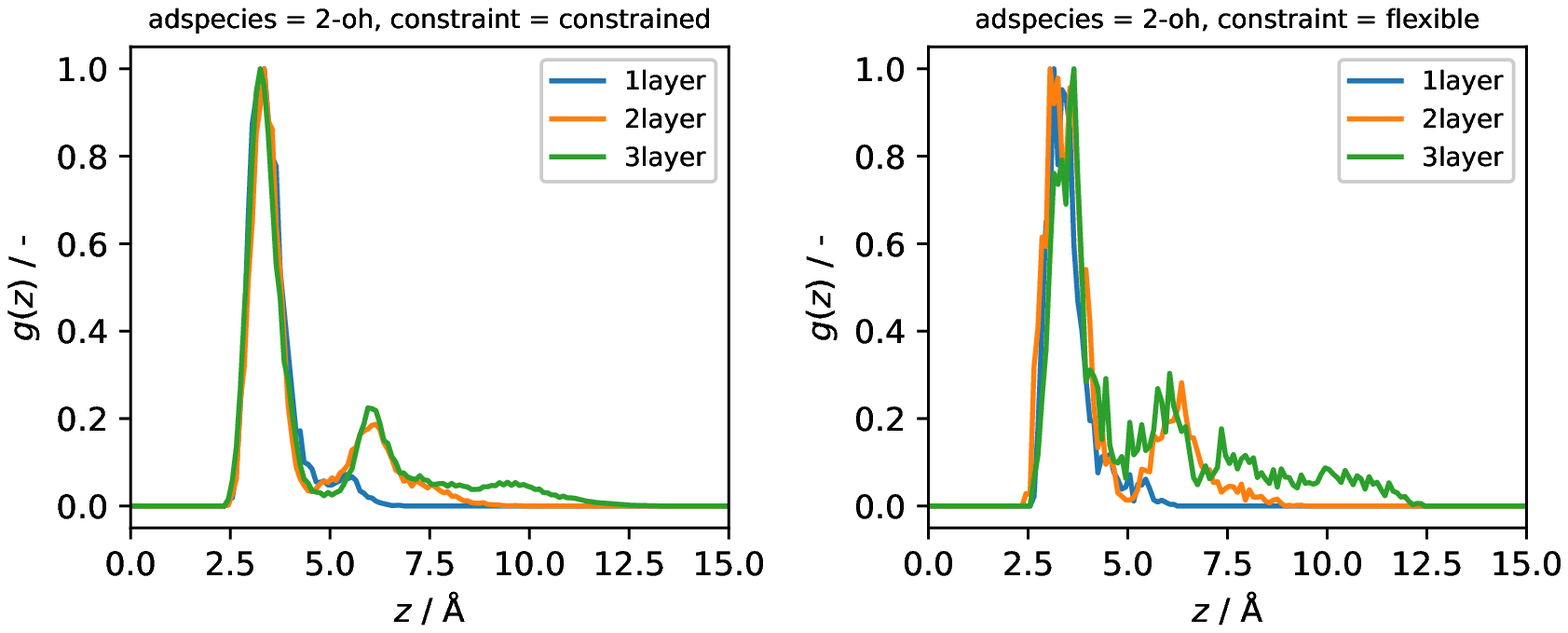}
    \includegraphics[width=0.49\linewidth]{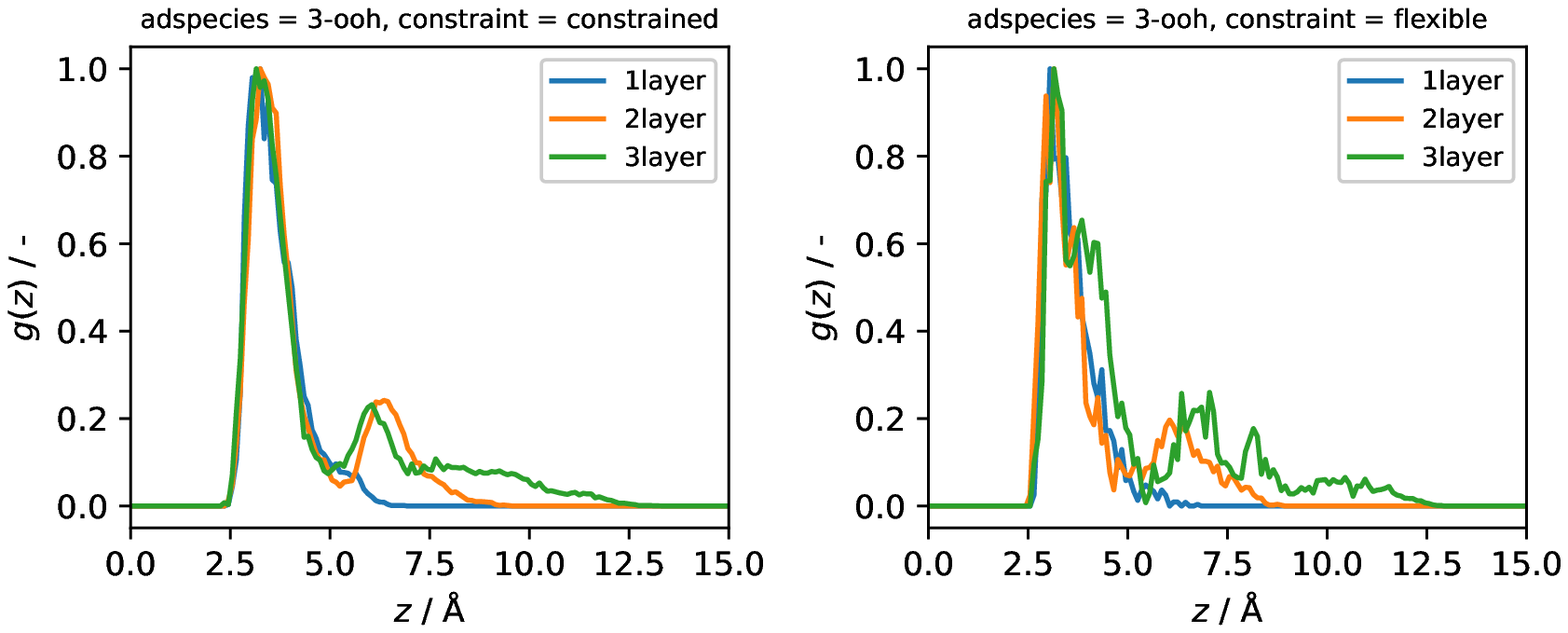}
    \caption{Distributions of O atoms in $z$ direction, $g(z)$, for the different systems in contact with 1--4 layers (8--32 molecules) of water: clean BG sheet model without adspecies, BG sheet with *O adatom, BG sheet with *OH admolecule, and BG sheet with *OOH admolecule. Results from the \textit{constrained MD} simulations (100 ps total) are always shown on the left, results from the \textit{flexible MD} (10 ps total) are shown on the right. The $g(z)$ is sampled every 100 fs in both cases. Distances are calculated between all water O atoms and the $x-y$ plane located inside the BG sheet model. Results were normalized so that the maximum $g(z)$ value in every distribution is 1.0 for better comparability.}
    \label{f:rdfs}
\end{figure}

Figure \ref{f:rdfs} shows results for the $z$ distribution of O atoms in the model systems. The distributions were obtained from the \textit{constrained MD} and the \textit{flexible MD} data sets. The distance pairs were calculated only from pairs that involved the surface as one of the partners; hence, the surface is located at $z = 0$ in the figures below and the distribution can be interpreted as the coordination of water atoms relative to the surface. The adspecies (*O, *OH, *OOH) are omitted from the analysis since the surface and adspecies were frozen during MD simulations and would show up as a sharp density peak higher than bands resulting from the water structure which is actually of interest.

Figure \ref{f:rdfs} shows that results from the \textit{constrained MD} data set are smoother due to the factor 10 longer sampling. However, constraining the water molecules also appears to remove some of the fine structure observed in the case of the \textit{flexible MD} data set; this is most apparent for the *OH and *OOH adspecies in contact with 3 layers of water molecules. This observation may hint towards the Rattle constraint changing the interaction with the surface and adspecies but more research is needed to make sure that this observation is not noise, \textit{i.e.}, the result of poor sampling statistics.

\subsection{Influence of the MD sampling frequency on the solvation stabilization energy results}

\begin{figure}[htbp]
    \centering
    \includegraphics[width=0.49\linewidth]{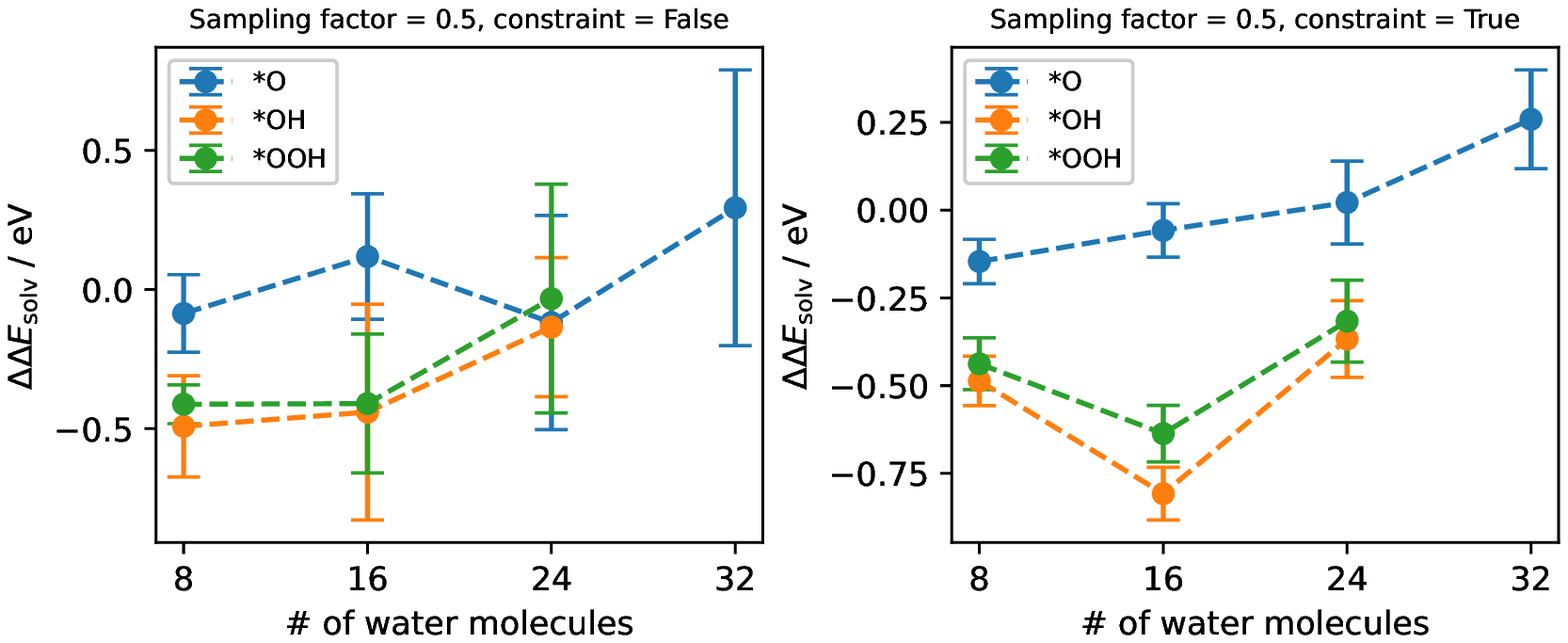}
    \includegraphics[width=0.49\linewidth]{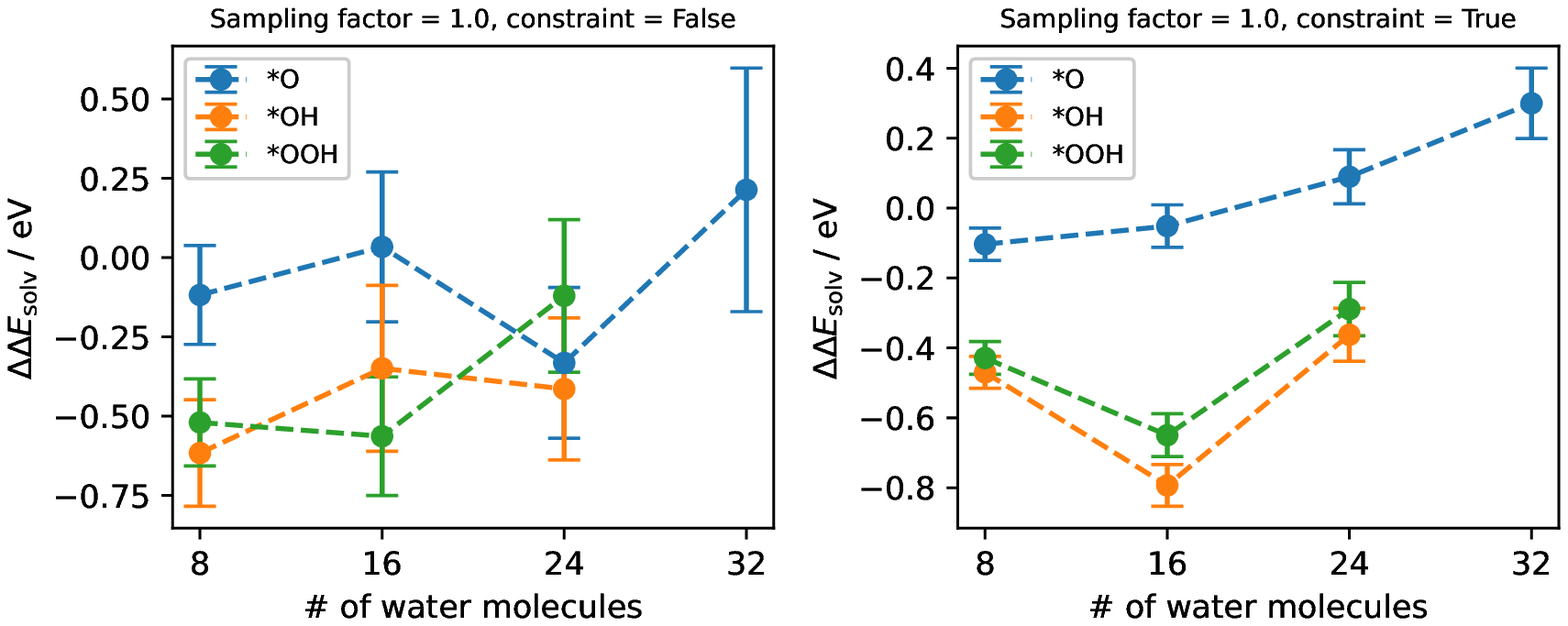}\\
    \includegraphics[width=0.49\linewidth]{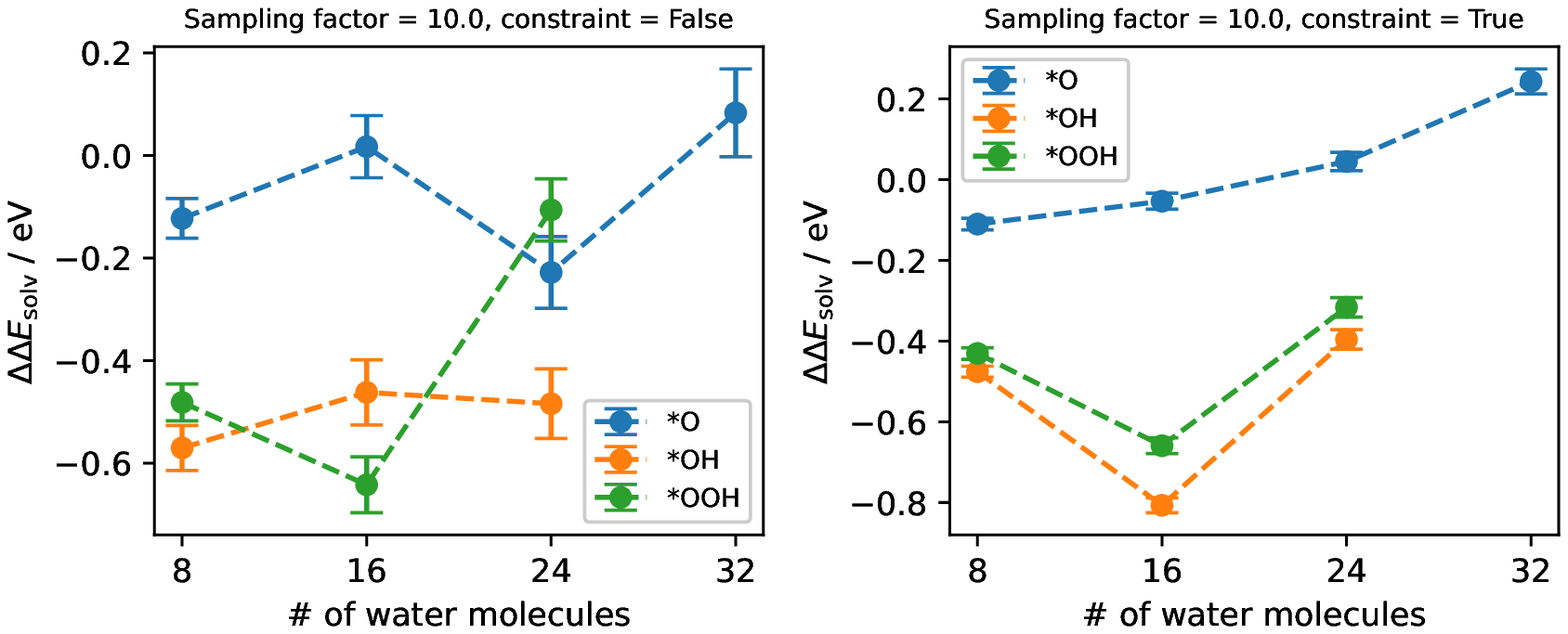}
    \includegraphics[width=0.49\linewidth]{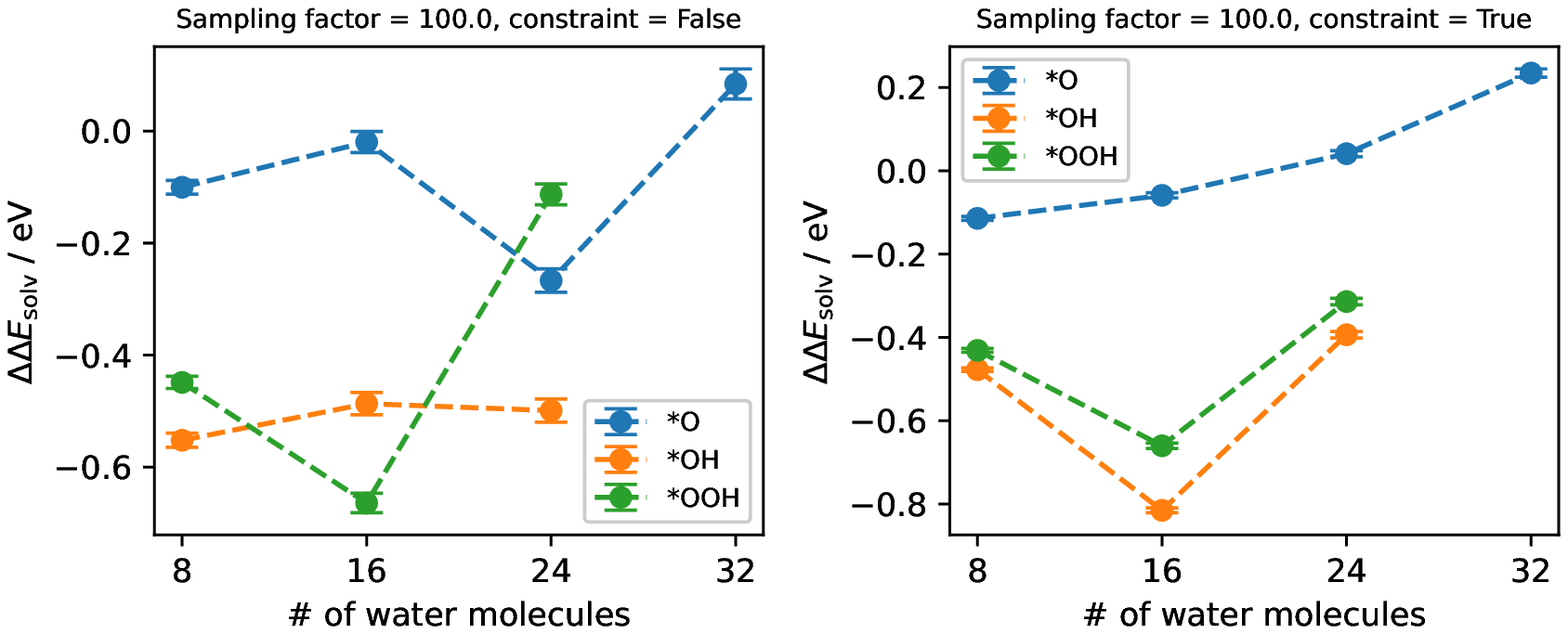}
    \caption{Influence of the MD sampling frequency on the $\Delta \Delta E_\mathrm{solv}$ results. The original sampling frequency shown in the main manuscript is 1 ps, indicated in this figure as sampling factor = 1. Shown here are additional sampling factors of 0.5, 10.0, and 100.0 which correspond to sampling frequencies of 2 ps, 100 fs, and 10 fs.}
    \label{f:sampling}
\end{figure}

Figure \ref{f:sampling} summarizes \DDEsolv\ results from the \textit{flexible MD} and \textit{constrained MD} data sets analogous to figure 4 in the main manuscript except that different intervals of sampling are tested (2 ps, 1.0 ps, 100 fs, and 10 fs represented by sampling factors 0.5, 1.0, 10.0, and 100.0, respectively). This test probes the robustness of the results against the sampling frequency and visualizes the impact that assuming more or less independent samples has on the obtained confidence intervals.

Figure \ref{f:sampling} shows that the average \DDEsolv\ results are robust against the sampling frequency. The only significant change is observed going from sampling factor 0.5 (2 ps) to factor 1.0 (1 ps) in the case of the \textit{flexible MD} data set. This change constitutes the the difference between 5 and 10 evaluated images for this data set. It can therefore be concluded that 10 independent images is the minimum number of images required to obtain a robust average \DDEsolv\ from this data set.

The choice of sampling frequency affects the size of the error bars significantly. This result highlights that it is important to chose the sampling frequency according to physical considerations (here: correlation time of water) since only checking for convergence of the average results can create a false sense of security from oversampling the data.

\subsection{Influence of the dipole and quadrupole correction on the results}

\begin{figure}[htbp]
    \centering
    \includegraphics[width=\linewidth]{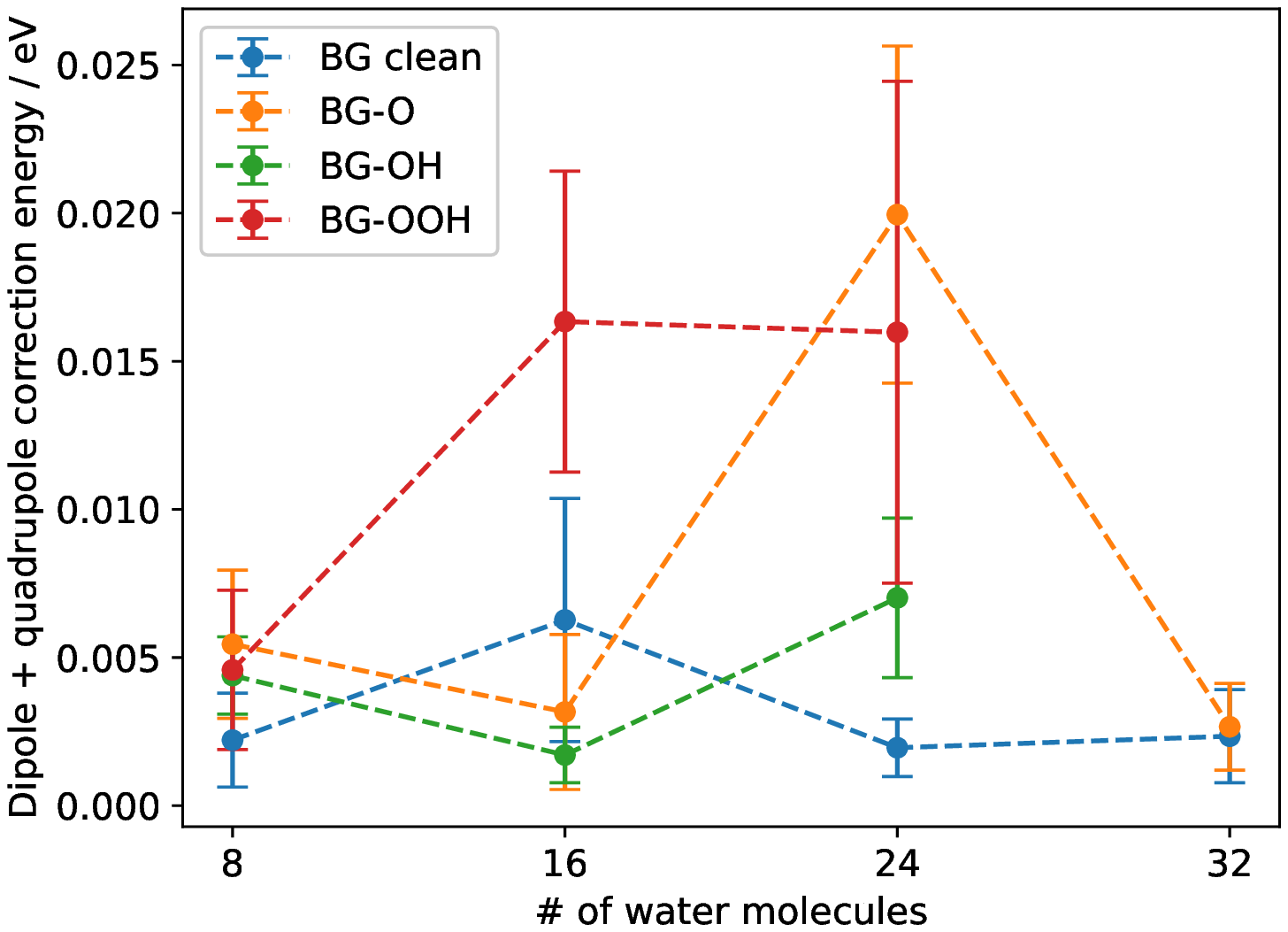}
    \caption{Average dipole and quadrupole energy correction values for the systems in the \textit{resampled} data set. While large fluctuations are noted in particular for the *OOH adspecies in contact with 24 and 16 water molecules and the *O adspecies in contact with 24 water molecules, the fluctuations remain below the limit of chemical accuracy (< 0.05 eV). The error bars indicate the two-sided 95 \% confidence interval.}
    \label{f:dipole}
\end{figure}

This test is performed to check if potentially erratic dipole and quadrupole correction values, which were empirically observed in this work to sometimes occur for no apparent reason, are causing the \DDEsolv\ results to be erratic. Figure \ref{f:dipole} visualizes the dipole and quadrupole energy correction results for the \textit{resampled} data set.

Results show that the average correction energy values are below 0.05 eV and therefore within chemical accuracy. However, the error bars for some systems are large, in particular for the *OOH adspecies in contact with 16 and 24 water molecules and the *O adspecies in contact with 24 molecules. This observation indicates that the correction can be somewhat erratic for certain systems and arrangements. Ultimately, the low absolute values for this correction are unlikely to impact results in a significant way.

\subsection{Influence of the dispersion correction on the results}

\begin{figure}[htbp]
    \centering
    \begin{minipage}[t]{0.49\linewidth}
        \vspace{0pt}
        \includegraphics[width=\linewidth]{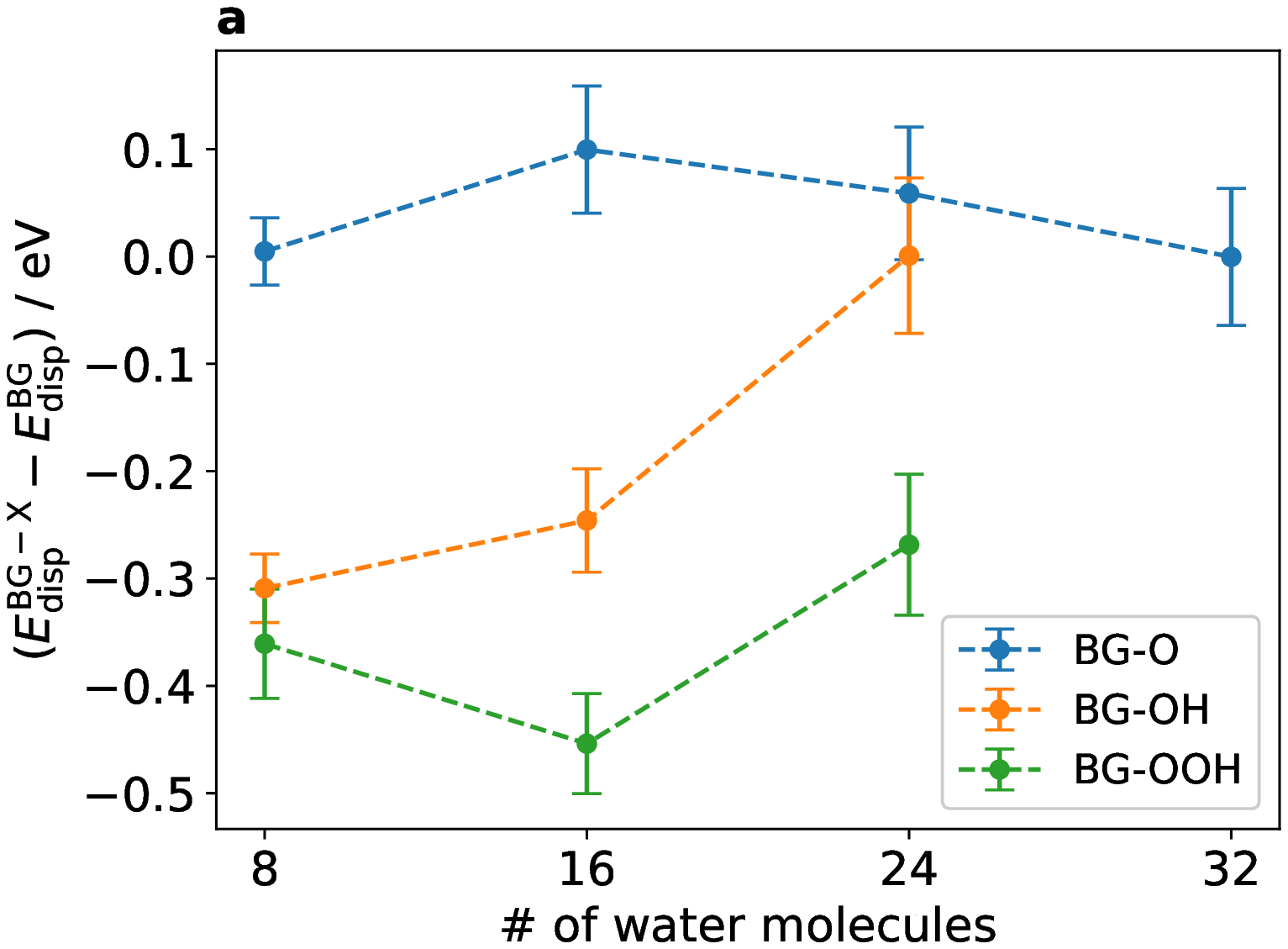}
    \end{minipage}
    \begin{minipage}[t]{0.49\linewidth}
        \vspace{0pt}
        \includegraphics[width=\linewidth]{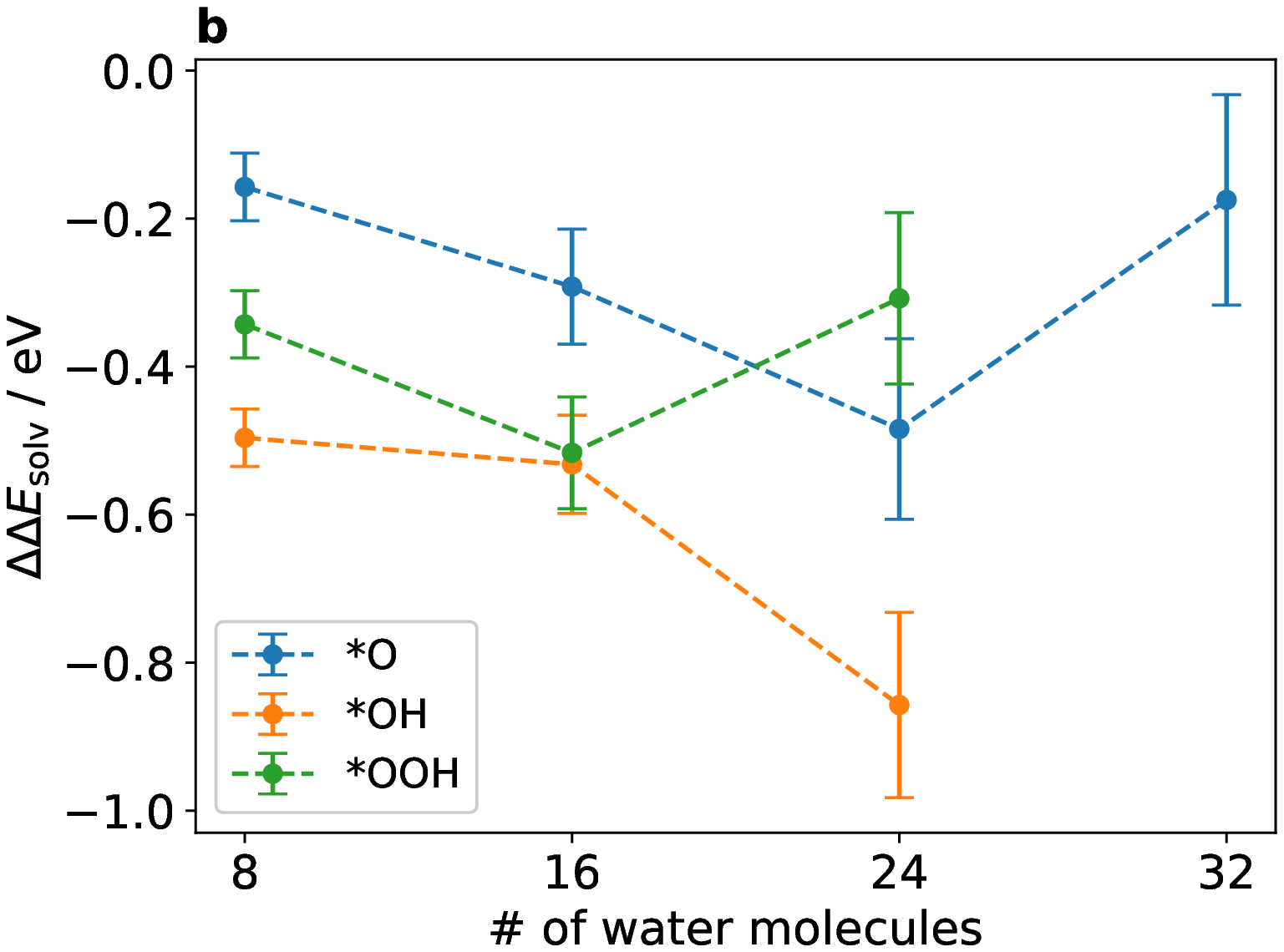}
    \end{minipage}

    \caption{Influence of the dispersion correction on the calculated \DDEsolv\ results. \textbf{a} Absolute dispersion correction energy for each tested system averaged over the data set of 20 images from the \textit{resampled} data set as outlined in the main article. Shown here is $\Delta E_\mathrm{disp} = E_\mathrm{disp}^\mathrm{BG-X} - E_\mathrm{disp}^\mathrm{BG}$, \textit{i.e.}, the difference of the dispersion contributions of the clean BG surface and the BG surface with an adspecies (X = O, OH, OOH). \textbf{b} \DDEsolv\ trends for the \textit{resampled} data set where $E_\mathrm{disp}$ has been deducted from the total energy values before calculating \DDEsolv. The error bars indicate the two-sided 95 \% confidence interval.}
    \label{f:dispersion}
\end{figure}

Similar to the dipole and quadrupole correction energy, the influence of the DFT-D3 dispersion correction on the obtained \DDEsolv\ results is tested. Figure \ref{f:dispersion}\textbf{a} shows the difference of the dispersion contributions of the clean BG surface and the BG surface with an adspecies (X = O, OH, OOH):
\begin{equation}
    \Delta E_\mathrm{disp} = E_\mathrm{disp}^\mathrm{BG-X} - E_\mathrm{disp}^\mathrm{BG}.
\end{equation}
Figure \ref{f:dispersion}\textbf{b} reproduces the \DDEsolv\ results for the \textit{resampled} data set shown in Figure 4 in the main manuscript but with the dispersion energy contribution removed from the total energy values before calculating \DDEsolv.

Results indicate that while the dispersion correction is significant in terms of absolute values, removing the $E_\mathrm{disp}$ does not stabilize the \DDEsolv\ trends either (figure \ref{f:dispersion}\textbf{b}). There is, however, an important caveat to this test: dispersion correction was included during the MD simulations from which the \textit{resampled} data set was generated and also during minimization calculations of the structures in the \textit{resampled} data set. Hence, this \textit{a posteriori} removal of the dispersion contribution can only be a rough indicator of its influence. The data sets would need to be reproduced completely without dispersion correction to conclusively rule out this parameter.

\subsection{Influence of the simulation box dimensions}

Table \ref{t:references} summarizes the total energy values of the reference systems without water molecules from the \textit{resampled} data set used to calculate \DDEsolv. The reference systems were generated from the parent systems that include water molecules by removing the latter. This step was necessary because the simulation box dimensions are different depending on how many water molecules are included. This table illustrates the differences introduced into the total energy by variable cell dimensions. The differences are < 0.01 eV in all cases and, even if the cell dimension had not been corrected for in this way, are unlikely to distort the simulation results in a meaningful way.

\begin{table}[htbp]
    \caption{Summary of total energy values of the reference systems without water molecules used to calculate \DDEsolv. The reference systems are from the \textit{resampled} data set.}
    \centering
    \renewcommand{\arraystretch}{1.5}
    \begin{tabular}{|r|c|c|}
    \hline
        System & \# of H$_2$O & $E_\mathrm{tot}$ / eV \\ \hline
        BG sheet (clean) & 8 & -290.425863 \\
         & 16 & -290.425863 \\
         & 24 & -290.429015 \\
         & 32 & -290.426769 \\ \hline
        BG-O* & 8 & -295.994912 \\
         & 16 & -295.994912 \\
         & 24 & -295.998699 \\
         & 32 & -295.996450 \\ \hline
        BG-OH* & 8 & -300.638397 \\
         & 16 & -300.638397 \\
         & 24 & -300.638397 \\ \hline
        BG-OOH* & 8 & -304.701104 \\
         & 8 & -304.701104 \\
         & 8 & -304.701104 \\ \hline
    \end{tabular}
    \label{t:references}
\end{table}

\subsection{Influence of energy-minimizing the non-solvated reference systems}

\begin{figure}[htbp]
    \centering
    \includegraphics[width=\linewidth]{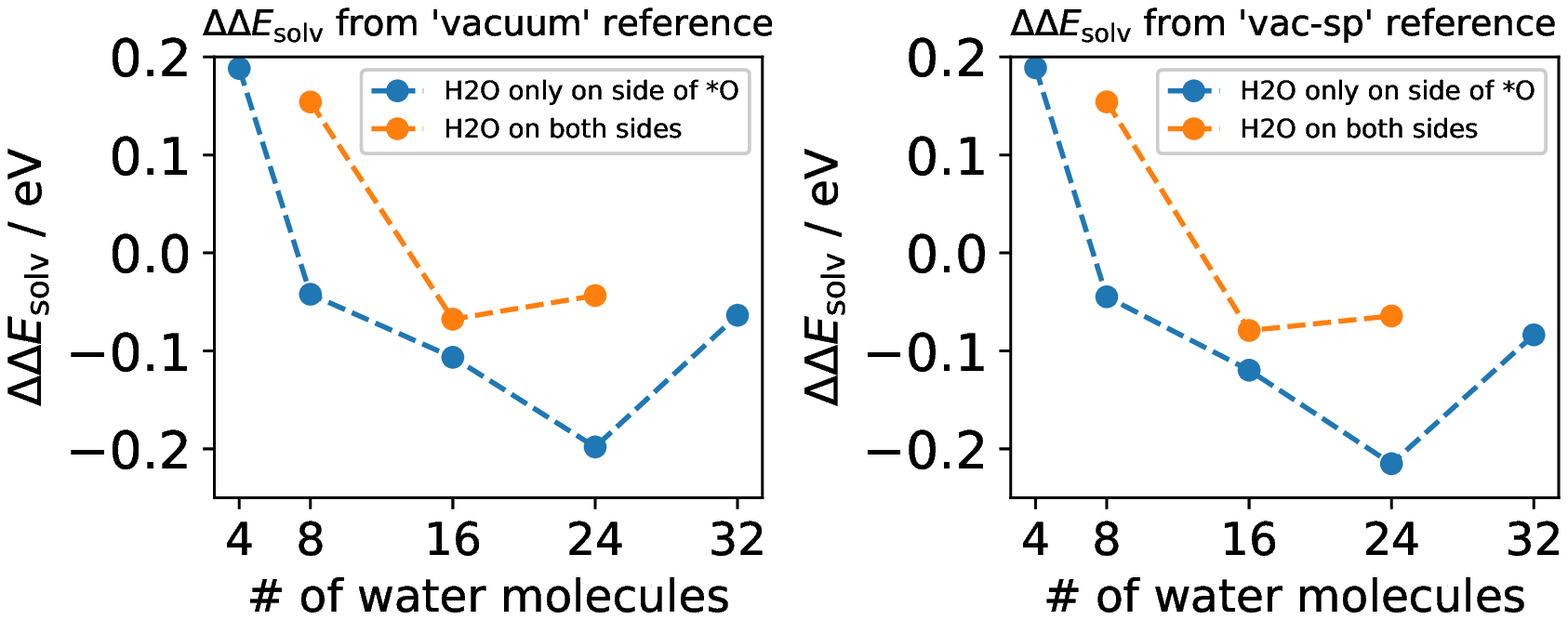}
    \caption{Exploring the influence of energy-minimizing the non-solvated reference systems. Left: non-solvated reference configurations were energy-minimized with respect to the atomic coordinates ('vacuum'). Right: non-solvated reference systems were not energy-minimized and energy values were obtained from single-point calculations ('vac-sp').}
    \label{f:minimization}
\end{figure}

Another potential source of inconsistency is the way that total energy values for the non-solvated reference systems are obtained. In the main article, the configurations of the non-solvated systems were obtained by removing the water molecules from the solvated systems and minimizing the resulting atomic configurations. However, with this approach, \DDEsolv\ does not only contain the interaction between the surface-adspecies system with the water molecules but also the rearrangement energy from the relaxation of the reference system. Figure \ref{f:minimization} explores for the \textit{one-shot minimization} data set whether \textit{not} minimizing the reference systems, \textit{i.e.}, obtaining the reference energy values from single-point calculations on the formerly-solvated systems where water molecules have been removed, makes a difference.

Results show that the differences are minimal at best. The relative trend does not change at all but absolute values are slightly more negative in the case of the single-point reference calculations compared to the minimized reference calculations for 16, 24, and 32 water molecules.

 \subsection{The solvation stabilization energy obtained only from the lowest-energy MD configurations}

 An article by Yu \textit{et al.} from 2011 details yet another way of obtaining \DDEsolv.\cite{Yu2011} The group performed MD simulations with NG model systems and the ORR adspecies in contact with explicit water molecules. They then picked the lowest-energy configuration generated in each MD simulation and performed an energy minimization calculation with respect to the atomic coordinates. The minimized systems were then used to calculate solvated free energy values.

\begin{figure}[htbp]
     \centering
     \includegraphics[width=0.8\linewidth]{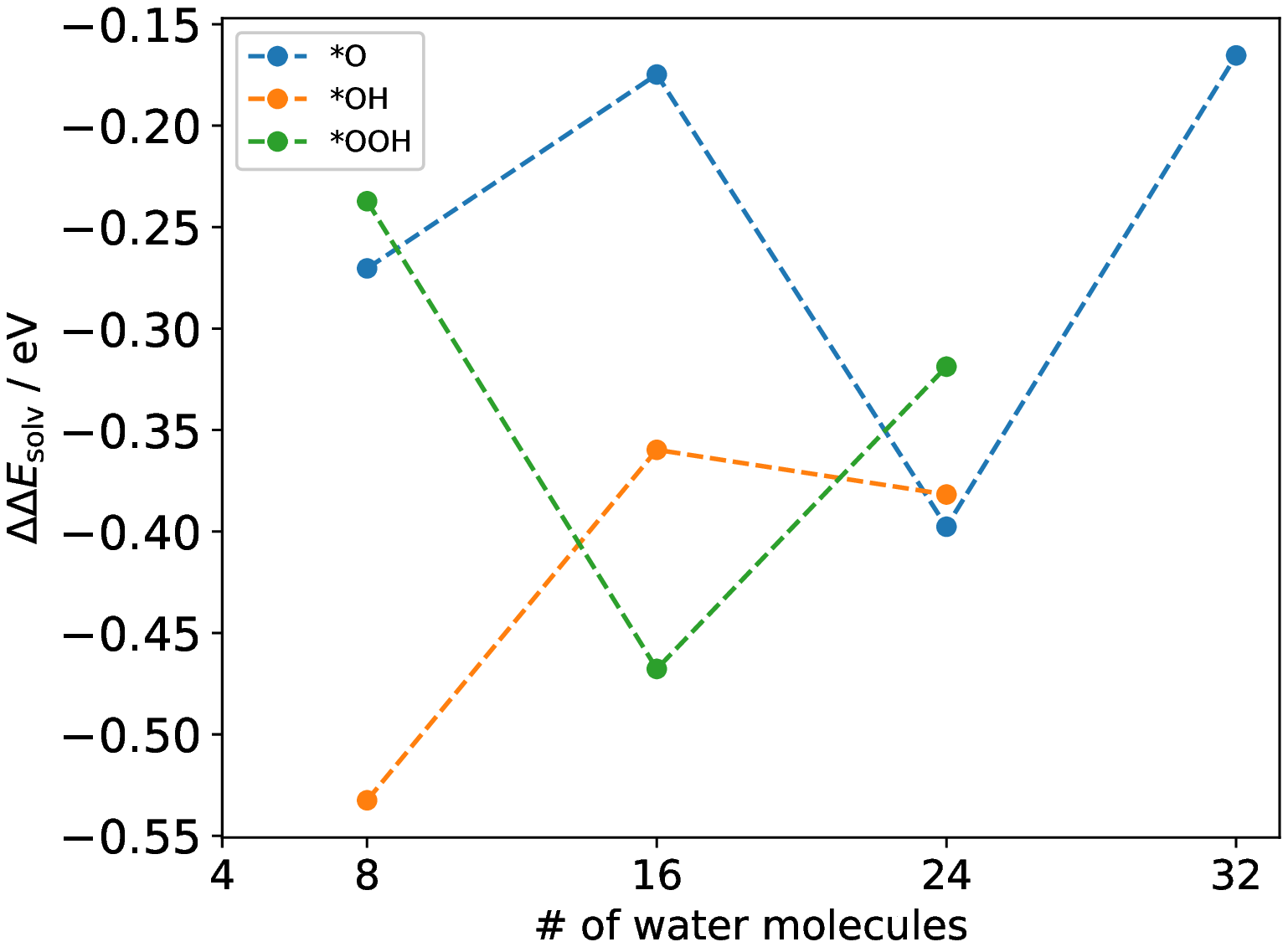}
     \caption{\DDEsolv\ obtained from only the lowest-energy configurations in the \textit{flexible MD} data set which were subsequently energy minimized.}
     \label{f:lowest}
 \end{figure}

 Figure \ref{f:lowest} applies this strategy to the \textit{flexible MD} data set.

 The closest comparison for this analysis are the results from the \textit{resampled} data set, see figure 4 in the main article. While the trends for *OH and *OOH in figure \ref{f:lowest} loosely resemble the trends for the \textit{resampled} data set, \DDEsolv\ for the *O adspecies is significantly more negative than with any other analysis strategy. It can be concluded that this approach not only did not resolve the erratic trends but likely further distorted the results because the close-to-ideal local configurations optimized in this case likely do not represent the average configurations of water molecules around the adspecies in real, finite-temperature systems.

 \subsection{Influence of freezing the BG sheet}

\begin{figure}[htbp]
   \includegraphics[width=\linewidth]{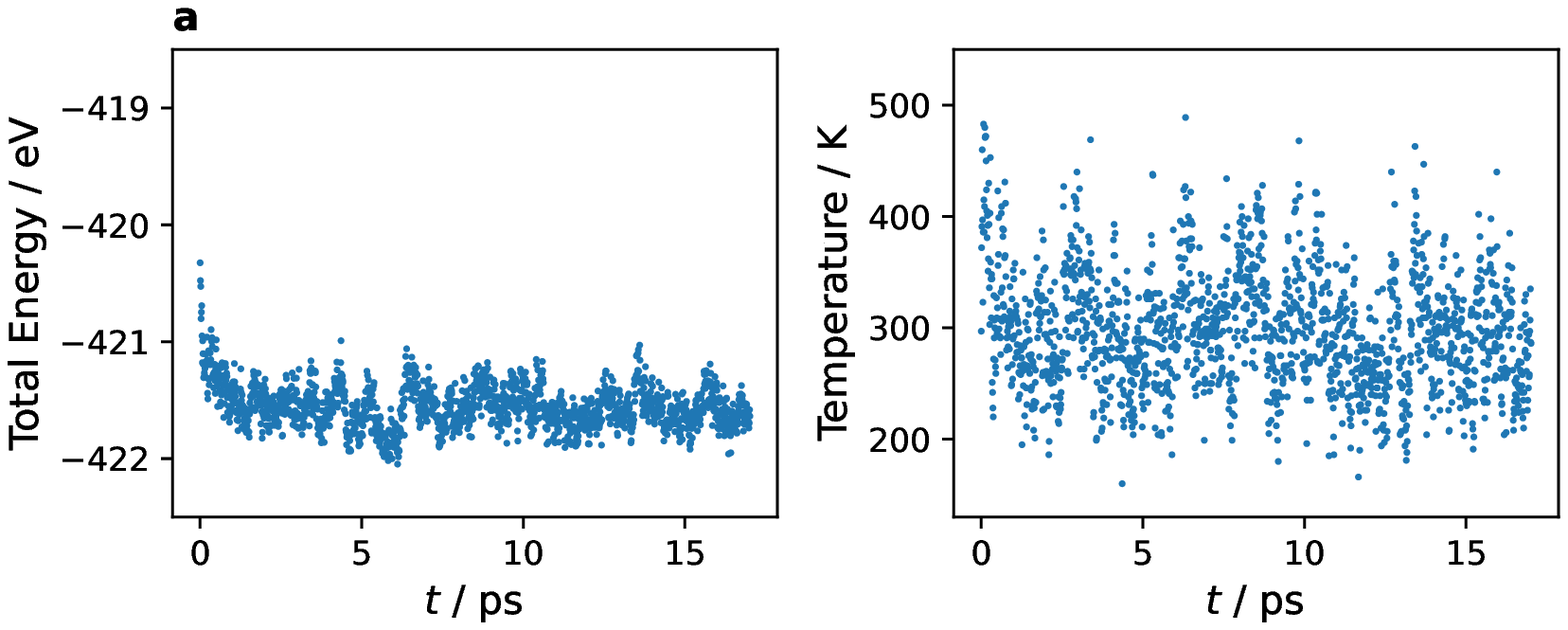}\\
   \includegraphics[width=\linewidth]{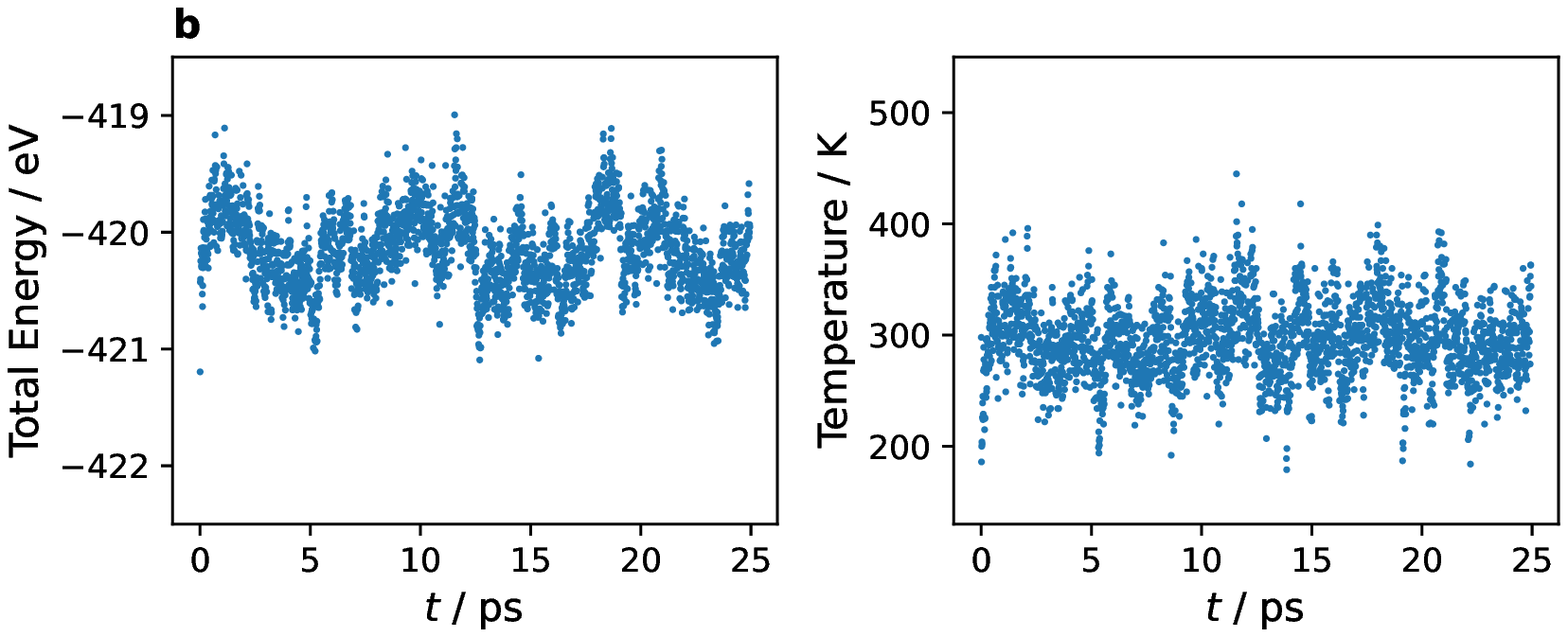}
   \caption{Comparison of $E_\mathrm{tot}$ \textit{vs.} $t$ and $T$ \textit{vs.} $t$ trends of \textit{flexible MD} simulations with the properly constrained BG sheet (\textbf{a}) and the unconstrained BG sheet (\textbf{b}).}
   \label{f:freezing1}
\end{figure}

\begin{figure}[htbp]
    \centering
    \includegraphics[width=0.8\linewidth]{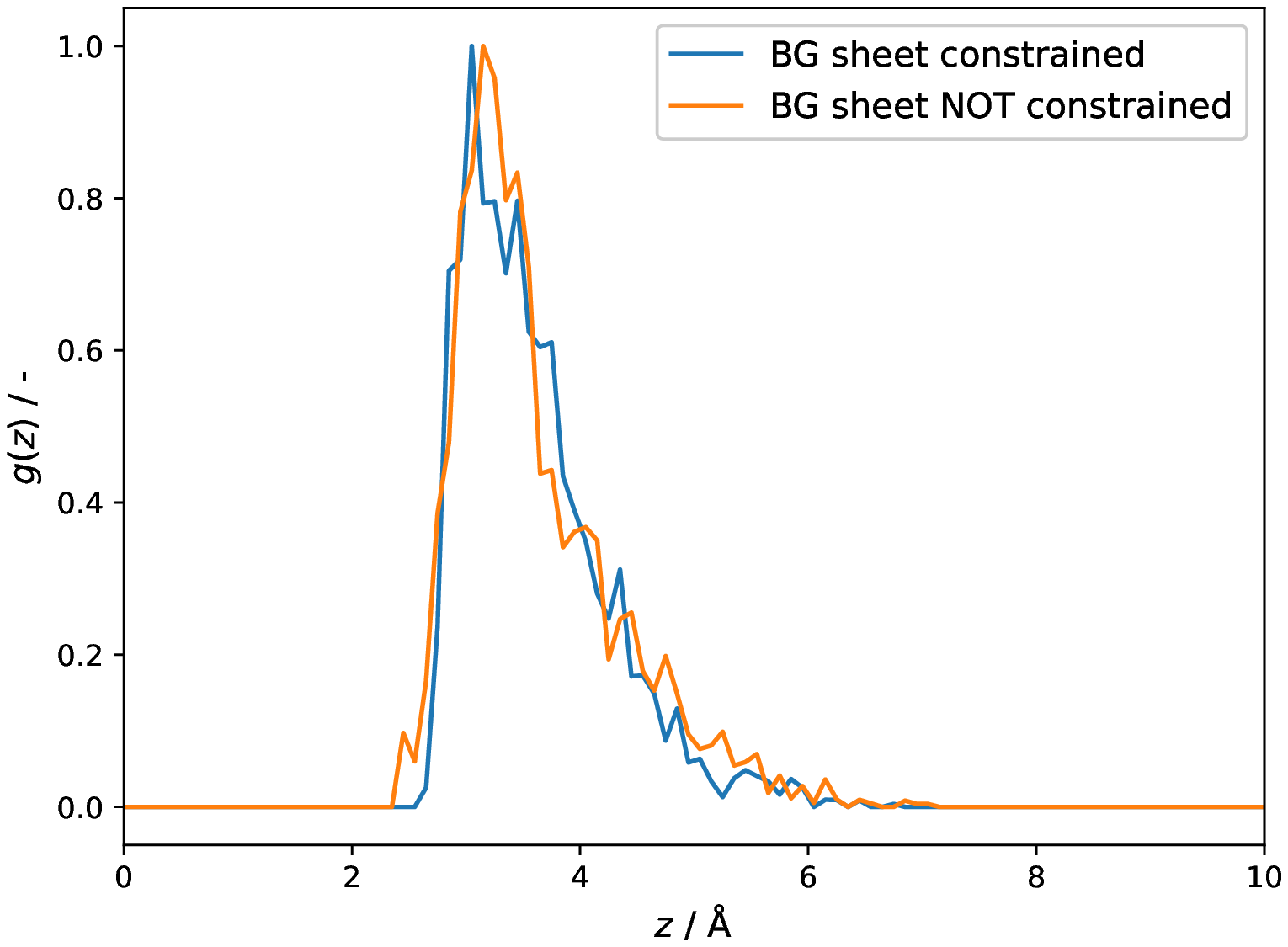}
    \caption{Comparison surface-O $g(z)$ distributions obtained from \textit{flexible MD} simulations using the constrained and non-constrained BG sheet.}
    \label{f:freezing2}
\end{figure}

The \textit{flexible MD} simulation of BG-OOH in contact with 8 (flexible) water molecules was accidentally performed with a non-constrained BG sheet. While the data shown in the main article was obtained with the correctly constrained model, this mistake allows to probe the influence of this geometry constraint. Figure \ref{f:freezing1} compares the $E_\mathrm{tot}$ \textit{vs.} $t$ and $T$ \textit{vs.} $t$ trends of \textit{flexible MD} simulations with the unconstrained BG sheet (\textbf{a}) and the properly constrained BG sheet (\textbf{b}). Figure \ref{f:freezing2} compares the surface-O $g(z)$ distributions obtained from \textit{flexible MD} simulations with these two models.

Figure \ref{f:freezing1} shows that the total energy and temperature fluctuations are significantly increased if the BG sheet with the adspecies is not constrained (\textbf{b}). This result may indicate that significantly more sampling would be required in the case of the non-constrained BG sheet to obtain a good estimate of \DDEsolv\ with a small-enough error bar. Figure \ref{f:freezing2} does not show any significant differences between the two systems, indicating that freezing the BG sheet and adspecies does not significantly change the interaction with the first layer of solvent molecules.

\end{document}